\documentclass[12pt]{article}
\usepackage{latexsym}
\usepackage{epsf}
\usepackage{hyperref}

\def\NON{\nonumber\\}

\def\a{\alpha}
\def\b{\beta}
\def\c{\chi}
\def\d{\delta}
\def\e{\epsilon}                
\def\g{\gamma}
\def\h{\eta}

\def\j{\psi}

\def\l{\lambda}
\def\m{\mu}

\def\o{\omega}
\def\p{\pi}                     
\def\th{\theta}                  
\def\r{\rho}                    
\def\s{\sigma}                  
\def\t{\tau}

\def\x{\xi}

\def\G{\Gamma}
\def\J{\Psi}

\def\S{\Sigma}

\def\cb{{\cal B}}
\def\cc{{\cal C}}
\def\cd{{\cal D}}

\def\cf{{\cal F}}
\def\cg{{\cal G}}
\def\ch{{\cal H}}   

\def\cl{{\cal L}}

\def\cn{{\cal N}}
\def\co{{\cal O}}
\def\cp{{\cal P}}

\def\car{{\cal R}}

\def\cu{{\cal U}}

\def\cbo{{\,\raise-.15ex\Sc [\,}}                       
\def\dg{^\dagger}                                     

\def\sl#1{\rlap{\hbox{$\mskip 1 mu /$}}#1}      
\def\vev#1{\Big\langle #1 \Big\rangle}           

\def\sbra#1{\left\langle #1\right|}             
\def\sket#1{\left| #1\right\rangle}             
\def\svev#1{\left\langle #1\right\rangle}       

\def\ddt#1{{\buildrel {\hbox{\LARGE .\kern-2pt.}} \over {#1}}}

\def\secteq#1{ \setcounter{equation}{0}
               \renewcommand{\theequation}{#1.\arabic{equation}} }

\def\APH#1{Ann. Phys. {\bf #1}}

\def\NPB#1{Nucl. Phys. {\bf B#1}}
\def\NPBP#1{Nucl. Phys. B (Proc. Suppl.) {\bf #1}}
\def\PLB#1{Phys. Lett. {\bf B#1}}
\def\PRD#1{Phys. Rev. {\bf D#1}}
\def\PR#1{Phys. Rev. {\bf #1}}
\def\PRL#1{Phys. Rev. Lett. {\bf #1}}
\def\PRP#1{Phys. Rep. {\bf #1}}
\def\RMP#1{Rev. Mod. Phys. {\bf #1}}

\def\sstyle{\scriptstyle}

\def\ie{\mbox{\it i.e.} }
\def\eg{\mbox{\it e.g.} }

\def\leqx{\,\raisebox{-1.0ex}{$\stackrel{\textstyle <}{\sim}$}\,}
\def\geqx{\,\raisebox{-1.0ex}{$\stackrel{\textstyle >}{\sim}$}\,}

\def\frac#1#2{ {\sstyle {#1\over #2} } }

\def\half{{1\over 2}}
\def\Re{{\rm Re\,}}

\thicklines                         
\def\bj{\overline\psi}

\def\eg{{\it e.g.\ }}
\def\ie{{\it i.e.}}
\def\cf{{\it cf.}}

\def\bj{\overline\psi}
\def\bl{\overline{l}}

\def\db{\partial^*}

\def\tG{\tilde\G}
\def\mres{m_{res}}
\def\gh{\hat\gamma_5}
\def\hT{\hat{T}}

\def\hatc{\hat{c}}

\begin{document}
\hyphenation{fer-mio-nic per-tur-ba-tive pa-ra-me-tri-za-tion
pa-ra-me-tri-zed a-nom-al-ous}

\begin{center}
\vspace{10mm}
{\large\bf Localization in Lattice QCD}
\\[12mm]
Maarten Golterman$^a$\ \ and \ \ Yigal Shamir$^b$
\\[8mm]
{\small\it
$^a$Department of Physics and Astronomy,
San Francisco State University\\
San Francisco, CA 94132, USA}\\
{\tt maarten@quark.sfsu.edu}
\\[5mm]
{\small\it $^b$School of Physics and Astronomy\\
Raymond and Beverly Sackler Faculty of Exact Sciences\\
Tel-Aviv University, Ramat~Aviv,~69978~ISRAEL}\\
{\tt shamir@post.tau.ac.il}
\\[10mm]
{ABSTRACT}
\\[2mm]
\end{center}

\begin{quotation}
In this paper, we examine the phase diagram of
quenched QCD with two flavors of Wilson fermions, proposing the
following microscopic picture.
The super-critical regions inside {\it and} outside the Aoki phase are
characterized by the existence of a density of
near-zero modes of the (hermitian) Wilson--Dirac operator, and thus
by a non-vanishing pion condensate.
Inside the Aoki phase, this density is built up from
extended near-zero modes, while outside the Aoki phase,
there is a non-vanishing density of exponentially localized
near-zero modes,
which occur in ``exceptional" gauge-field configurations.
Nevertheless, no
Goldstone excitations appear outside the Aoki phase, and the
existence of Goldstone excitations may therefore be used to define
the Aoki phase in both the quenched and unquenched theories.
We show that the density of localized near-zero modes gives rise to
a {\it divergent} pion two-point function, thus providing an alternative
mechanism for satisfying the relevant Ward identity in the presence
of a non-zero order parameter.
This divergence occurs when we take a ``twisted" quark mass to zero,
and we conclude that quenched QCD with Wilson fermions is
well-defined {\it only} with a non-vanishing twisted mass.
We show that this peculiar behavior of the near-zero-mode density
is special to the quenched theory by demonstrating
that this density vanishes in the unquenched theory outside the Aoki phase.
We discuss the implications for domain-wall and
overlap fermions constructed from a Wilson--Dirac kernel.
We argue that both methods work outside the Aoki phase,
but fail inside because of problems with locality and/or
chiral symmetry, in both the quenched and unquenched theories.
\end{quotation}

\newpage
\vspace{5ex}
\noindent {\large\bf 1.~Introduction}
\secteq{1}
\vspace{3ex}

Wilson fermions play a prominent role in Lattice QCD
\cite{wilson}.  They
(and their improved versions) are widely used in
numerical calculations of hadronic quantities.  In addition,
they are at the heart of the construction of lattice Dirac
operators with domain-wall \cite{dwf,fs} or overlap \cite{oovlp,ovlp}
fermions.  All these
fermion methods based on the Wilson--Dirac operator preserve
all of the flavor symmetry, but not ordinary chiral symmetry.  For
Wilson fermions, a tuning of the bare fermion masses is needed
to restore chiral symmetry in the continuum limit, while
domain-wall fermions (with infinite extent in the fifth
dimension) and overlap fermions possess a modified version
of chiral symmetry with essentially the same algebraic properties as
the chiral symmetry of the continuum theory \cite{mlsymm}.
This lattice chiral symmetry reduces to that of the continuum
theory in the continuum limit.

In this paper, we will discuss the phase diagram
in the gauge-coupling, quark-mass plane for two degenerate
Wilson fermions, for both the quenched and unquenched theories.
We will be interested in correlation functions constructed
from the inverse of the two-flavor (hermitian) Wilson--Dirac
operator, evaluated on an equilibrium ensemble of gauge-field
configurations.  The difference between the quenched and
unquenched cases is that only in the unquenched case the
fermion determinant (which is positive for two degenerate
flavors) will be part of the Boltzmann weight used to generate
the ensemble, whereas in the quenched case the Boltzmann
weight is obtained from a local pure-gauge action.
The quenched theory can be understood as a euclidean path
integral with two physical quarks, and two ``ghost" (bosonic)
quarks with the same mass, whose role it is to cancel the
physical-quark determinant \cite{morel,bg}.

Long ago, a conjecture was made for this phase diagram \cite{aoki},
which is shown schematically in Fig.~1, where $g_0$ is the bare
coupling and $m_0$ is the bare quark mass.  We will be concerned
with the region
$-8\le am_0\le 0$ because only in that region can the Wilson--Dirac
operator have any zero or near-zero eigenvalues.
\begin{figure}[thb]
\vspace*{0.4cm}
\centerline{
\epsfxsize=9.0cm
\put(-120,0){\epsfbox{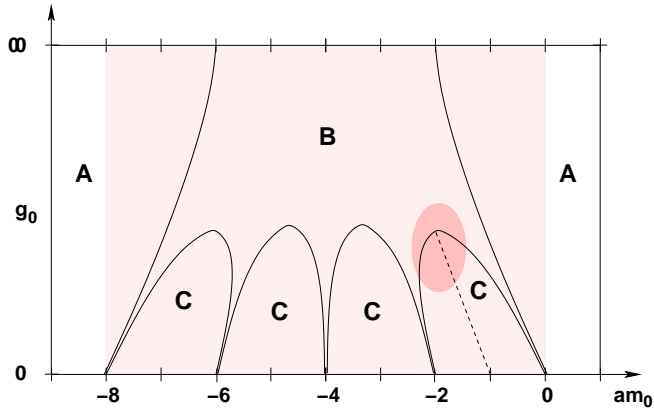}}
}
\vspace*{0.4cm}
\caption{ \noindent {\it
A representation of the phase diagram proposed by Aoki for
two-flavor QCD with Wilson fermions and standard plaquette action.
The solid lines mark phase transitions (all believed to be continuous).
Phase B is the Aoki phase,
defined by the existence of a parity- and flavor-breaking pionic
condensate.
Its ``trade mark'' are the five ``fingers'' reaching to the
critical points on the $g_0=0$ axis. Phases A and C have no condensate.
The lightly shaded area marks the super-critical region where near-zero
modes may occur. The lines $m_0=0$ and $am_0=-8$,
that define the boundaries of the super-critical region,
appear to have no special dynamical significance:
the A phases extend on both sides of them.
The quenched phase diagram is discussed in the main text.
The darker shaded area is roughly the area where domain-wall fermion
simulations have been carried out.
The dashed line represents a possible trajectory for taking the continuum
limit with domain-wall or overlap fermions (see Sect.~7).
}}
\end{figure}
Spontaneous symmetry breaking (SSB) occurs as a consequence of
a non-zero density of such modes \cite{bankscasher}, and the
interesting part of the phase diagram is thus confined to this
region (the ``super-critical" region).  The usual continuum limit
corresponds to the critical point at $g_0=0$, $am_0=0$.
Other continuum limits are obtained for any value of $am_0$
by taking $g_0\to 0$, but generically all quarks stay massive
(\ie\ have a mass of order of the cutoff).  Only at $-am_0=0,2,4,6,8$
do massless quarks show up in the continuum limit, respectively
2, 8, 12, 8, 2 of them (in our two-flavor theory).  The number
of quarks is determined by the number of momenta (or ``corners'')
in the Brillouin zone where the free Wilson--Dirac operator has a zero.

According to this conjecture,
in regions A and C no SSB takes place, while region B, the Aoki phase,
is defined
by the existence of a pionic condensate, which breaks parity and
flavor symmetry.\footnote{
  Note that all chiral symmetries are explicitly broken at non-zero
  lattice spacing, and thus play no role in determining the phase diagram.
}
Let us briefly review the evidence for the (unquenched) Aoki phase diagram.
First, a simple heuristic argument suggests that a pionic
condensate must exist in some part of the phase diagram \cite{aoki}.
Start with some $m_0 \ge 0$ and examine what happens as $m_0$
is decreased. For a non-zero lattice spacing $a$
(\ie\ $g_0>0$) there is no chiral symmetry.
Instead of the continuum relation $m_\p^2 \propto |m_q|$
(here $m_\p$ and $m_q$ are respectively the
pion and quark masses) we have that
$m_\p^2 \propto m_0 - m'$, where the fact that $m'=m'(g_0)$ does not
vanish is a consequence of the breaking of chiral symmetry on the lattice.
For $m_0>m'$ the pions are massive, becoming massless at the
critical line $m_0=m'(g_0)$.  For $m_0<m'$, $m_\p^2$ would
go negative, signalling the breaking of a symmetry.  A pionic condensate
forms in some direction in flavor space, and the line $m_0=m'(g_0)$
determines the location of a second-order phase transition.
The corresponding pion becomes massive again for $m_0<m'$,
while the other two pions become Goldstone bosons associated with
the spontaneous breaking of the SU(2) flavor symmetry (``isospin'')
down to a U(1) symmetry. Since the condensate is pionic, it breaks
parity symmetry as well.  Microscopically, the condensate
arises from near-zero modes of the Wilson--Dirac operator
\cite{bankscasher}, and can thus only occur in the region
$-8<am_0<0$.

This argument does not provide much information on the detailed
form of the Aoki phase.  Additional analytical
evidence comes from several sources.
The location of the critical points along the line
$g_0=0$ is obtained from weak-coupling perturbation theory,
which, however, gives no information on the existence of a condensate.
The existence of a condensate was discussed in the context
of the pion effective action in Refs.~\cite{creutz,shsi}.
Ref.~\cite{shsi} showed that, {\it if} the Aoki phase
extends all the way to $g_0=0$, it does so as indicated
by the ``finger" structure\footnote{
  A similar phase diagram is known to exist in the Gross-Neveu model
  \cite{aoki,KeSe}.
}
in Fig.~1, with the width of
these fingers proportional to $(a\Lambda_{QCD})^3$, where
$\Lambda_{QCD}$ is the QCD scale.
In the strong-coupling limit, the location of the
two critical points was first found in Ref.~\cite{kawamotosmit},
and the nature of the phase with broken symmetry
was clarified in Ref.~\cite{aoki}.

The phase diagram was studied numerically
in both quenched~\cite{Aokiq} and full~\cite{Aokif} QCD.
The numerical results provide evidence
that the five critical points at $g_0=0$ are indeed continuously
connected
to the ``main body'' of the Aoki phase at large coupling.
In the quenched theory,
evidence for all ten critical lines at $g_0=1$ ($\b=6.0$) was found in
the last paper of Ref.~\cite{Aokiq}.
In the present
paper, in which we address the mechanism responsible for the
existence of an Aoki phase, the detailed form of the phase
diagram is not important.  For the sake of argument, we will assume
that the very plausible situation depicted in Fig.~1 describes
the actual phase diagram.  For a discussion expressing some doubts,
see Ref.~\cite{EW}.

A difficulty with the quenched simulations~\cite{Aokiq} is that
exceptional configurations were discarded.  Exceptional
configurations~\cite{SV,excep} will play an important role below.
Keeping this subtlety in mind, no light excitations
were found in the quenched A and C phases, except close to the
Aoki (B) phase boundary
(and, in particular, close to the five critical points at $g_0=0$).

While this picture appears to be rather satisfactory, there
exists some evidence that, in the {\it quenched} case, seems to
disagree with the diagram of Fig.~1.  Numerical studies of
the near-zero modes of the (hermitian) Wilson--Dirac operator
indicate that, for $g_0>0$,
a non-zero density of near-zero modes {\it always} occurs
in the quenched theory, {\it anywhere} in the super-critical region
\cite{scri,jansenetal}
(namely, in regions B and C, and the super-critical part of region A).
Through the Banks--Casher relation, this would imply that the
pionic condensate vanishes nowhere in this region, and SSB
takes place everywhere, thus contradicting the phase
structure sketched in Fig.~1.  Moreover, one would be inclined to
expect Goldstone excitations everywhere in the super-critical region,
in which case Fig.~1 would be
completely wrong for the quenched theory; it would not even
serve as a guide to the long-range physics.  We note however,
that this is in conflict
with what is known from the numerical studies of Ref.~\cite{Aokiq}
mentioned above,
as well as with an analytical study \cite{GSS}, in which it
is argued that the effective-field theory analysis of
Ref.~\cite{shsi} is also valid for the quenched case.

Another clue comes from a study of a special
class of zero modes by Berruto, Narayanan and Neuberger
(which we will refer to as BNN hereafter) \cite{BNN}.
They showed that, for choices of $m_0$ in the super-critical
region, the Wilson-Dirac operator has exact zero modes for
very smooth gauge fields with one dislocation, which is
contained in a small hypercube with a linear size
of a few lattice spacings (later we give a more precise
description of their results).  These zero modes are
exponentially localized, and their existence characterizes
these ``BNN" gauge-field configurations as exceptional configurations.
They further argued that in large volume one would expect
configurations with a ``dilute gas" of such dislocations,
because of their highly localized nature.  This class of
configurations could then contribute to a
non-vanishing density of near-zero modes of the Wilson--Dirac
operator for the quenched theory, where such configurations
are not suppressed by the fermion determinant.  This study
thus lends analytical support to the numerical results reported
in Refs.~\cite{scri,jansenetal}.

In this paper, we will propose a resolution to this apparently
paradoxical situation.  First, we prove that, if there is a
non-vanishing density of exponentially localized near-zero modes,
there exists a different mechanism for saturating the Ward
identity involving the order parameter for SSB.
The orientation of an order parameter is determined
by a small external ``magnetic field,''
here provided by a so-called twisted-mass term \cite{aoki,tm}.
We argue that under the above circumstances,
no Goldstone pole need appear, but instead
that the two-point function of the would-be Goldstone-pion field
{\it diverges} in the limit of vanishing twisted mass,
even if the momentum does {\it not} vanish.\footnote{
  While completing this paper, we became aware of the fact that a similar
  observation has been made some time ago in condensed matter theory \cite{ks}.
}
This divergence already occurs in finite volume,
and we will argue that it persists
in the infinite-volume limit of the quenched theory.

We then invoke a concept from condensed-matter physics in order
to come up with a conjecture for the microscopic picture underlying the
quenched phase diagram.  The motivation is to understand what happens
in the case that we have a dense, rather than dilute, gas of
dislocations.  For this purpose, we interpret the square of the hermitian
Wilson--Dirac operator as the hamiltonian of a five-dimensional
theory, and thus its (positive) eigenvalues as energy eigenvalues.
The ensemble of gauge fields on which these eigenvalues are computed
act as a random potential, and a non-zero measure subset of these
fields can be viewed as random distributions of highly localized
scattering centers.  The eigenstates describe the possible states
of a (four-plus-one-dimensional) ``electron" in this background.  Note that
the fact that we are concerned with the quenched theory here makes
the simplicity of this picture more compelling, because of the absence
of any feedback of the fermions on the gauge fields.

In general, there will exist a band of extended
eigenstates above a certain energy, denoted $\l^2_c$, whereas
eigenstates below this energy will be localized exponentially.\footnote{
  In a less detailed form, this physical picture has previously been considered
  in Ref.~\cite{aokidw}.
}
While attractive-potential scattering centers
will tend to localize the electron
as in a bound state of an isolated scattering center,
a large-enough density of them will make it
possible for the electron to travel throughout the lattice by tunneling,
with the corresponding state becoming extended.  The outcome will
depend on the energy of the state
and on the density and other properties of the scattering centers.
The (average) localization range of the states
will increase as the energy
is increased towards $\l^2_c$,
becoming infinite at $\l^2_c$, which in condensed-matter physics is referred
to as the ``mobility edge"
\cite{lcl} (for reviews see Refs.~\cite{DJT,EF,KK}).

The value of the mobility edge is a dynamical issue,
which depends on the ensemble of gauge fields,
and thus on the location in the phase diagram. We conjecture that the
Aoki phase is precisely that region of the phase diagram in which
the mobility edge is equal to zero.  Outside the Aoki phase, it is
larger than zero, and consequently, all near-zero modes are
exponentially localized.  When one moves closer to the phase transition,
the mobility edge comes down continuously, until it vanishes and the
Aoki phase is entered.

We complete our discussion of the physical picture
by invoking our results on the
two different mechanisms by which the Ward identity can be saturated
in the presence of a non-zero condensate.
When the mobility edge is zero, only extended near-zero modes
contribute to the spectral density,
and the Ward identity predicts the existence
of Goldstone excitations that dominate the long-range physics of our theory;
when the mobility edge is larger than zero,
the condensate (and, hence, the order parameter) is produced by exponentially
localized near-zero modes, the pion two-point function is diverging,
and, we conjecture, no Goldstone excitations occur.
Thus the region with Goldstone bosons
coincides with the region where the mobility edge is equal to zero,
and the existence of Goldstone bosons can be used as a {\it definition}
of the Aoki phase in both unquenched and quenched QCD.

This picture of what the quenched phase diagram looks like begs
the question as to whether it is consistent with unquenched
QCD.  In the unquenched case, the Boltzmann weight is modified
by the fermion determinant, which tends to suppress the
entropy of gauge-field configurations with near-zero modes.
We show that the divergence in the pion two-point function
cannot occur in the unquenched case.  Therefore, indeed no
near-zero-mode density can build up unless the resulting condensate
is accompanied by Goldstone bosons (in which case we are
inside the Aoki phase),
and exponentially localized near-zero modes do not play the same role
in unquenched QCD as in quenched QCD.

The scale of the typical localization length of the near-zero modes
is set by the lattice spacing.
The existence of a non-zero condensate outside the Aoki phase
is thus a short-distance artifact.
One expects {\it no} long-range physics (such as the existence
of Goldstone excitations) as a consequence of localized
near-zero modes, unless their average localization length becomes
so large that they behave collectively.
According to our conjecture, this happens precisely when the
mobility edge comes down to zero, and, just as in unquenched QCD,
only extended near-zero modes contribute to the spectral density.
Therefore, an effective-lagrangian analysis
in terms of the long-range effective degrees of freedom,
as carried out in the unquenched case \cite{shsi},
should also make sense in the quenched case.
Both inside and close to the boundaries of the Aoki phase,
the effective lagrangian should provide a valid description
of the long-range physics.
Indeed, such an analysis has been carried out
in the quenched theory \cite{GSS}, and leads to
conclusions very similar to those obtained in the unquenched case.

Thus far, our results deal with the case of QCD with Wilson
fermions.  However, there are important consequences for
lattice QCD with domain-wall or overlap fermions constructed
from the Wilson--Dirac operator.
In short, we claim that domain-wall and overlap fermions
can only be used {\it well outside} the Aoki phase.
For domain-wall fermions, the hermitian Wilson--Dirac operator
is closely related to (the logarithm of) the transfer matrix
that hops the fermions in the fifth dimension.
A density of near-zero modes of the hermitian Wilson--Dirac operator
implies the existence of long-range correlations in the fifth direction.
This threatens the decoupling of the left-handed and the right-handed quarks
that live on opposite ends of the five-dimensional world,
and, hence, the restoration of chiral symmetry in the limit on an
infinite fifth dimension.

The crucial issue is the value of the mobility edge of
the Wilson--Dirac operator.
Well outside the Aoki phase the near-zero modes are
all exponentially localized, with a scale set by the lattice spacing.
As a result, their contribution to observables
vanishes when the four-dimensional separation is set by a physical scale
(without having to tune any parameter).
As for extended modes of this operator, they cannot mediate long-range
correlations in the fifth dimension, and thus they do not
obstruct the recovery of chiral symmetry either.
In contrast, inside the Aoki phase the near-zero modes are extended,
and they mediate long-range correlations in {\it all} five directions.
This strongly suggests that inside the Aoki phase
the resulting four-dimensional effective theory
either is completely non-local, or, at best, contains
long-range degrees of freedom different from the desired ones.
This physical picture is valid for any value of the
lattice spacing in the fifth dimension $a_5$,
and thus the same conclusion applies to overlap fermions
which corresponds to the special case $a_5\to 0$.

This paper is organized as follows.  We define our theory,
introduce the twisted quark mass, and derive the relevant Ward
identity in Sect.~2, where we also briefly review the relation
between the condensate and the near-zero-mode density.
In Sect.~3 we derive a spectral representation for the
pion two-point function.  We observe that in the super-critical
quenched theory this two-point function always diverges in finite volume,
in the limit of a vanishing twisted mass.
We thus provide an alternative mechanism for
saturating the Ward identity in the presence of a non-vanishing condensate.
In Sect.~4, we start with defining localized
near-zero modes more precisely. We show that a non-zero
density of them will dominate the divergence of the pion two-point function
in large volume, and, likely,
will produce the divergence in the infinite-volume limit as well.
We then give a qualitative but detailed discussion
of the mobility edge, and its role in determining which
way the Ward identity is satisfied.  This leads to a
comprehensive picture of the quenched phase diagram,
and in particular of the distinction of the super-critical regions inside
and outside the Aoki phase.  In Sect.~5 we discuss briefly
how the picture changes if we reintroduce the fermion determinant,
\ie\ if we unquench the theory.  We then discuss the implications
of our analysis for domain-wall and overlap fermions
in Sect.~6, and finally summarize our conclusions in Sect.~7.
In Appendix~A we review Anderson's definition of localization
as the (partial) absence of diffusion \cite{lcl}, and its relation
to our definition in Sect.~4.
Some technical details are relegated to Appendix~B.

\vspace{5ex}
\noindent {\large\bf 2.~Twisted-mass QCD,
the Ward identity, and the Aoki condensate}
\secteq{2}
\vspace{3ex}

The definition of the Wilson--Dirac operator is
\begin{eqnarray}
\label{DW}
   D(m_0)& = & {1\over a}\left(\begin{array}{cc}
      -(W + a m_0)    & C     \\
      -C\dg  & -(W + a m_0)
       \end{array}\right)\,,  \NON\\
   C_{xy} & = & \half \sum_\m \left[\d_{x+\hat\m,y} U_{x\m}
    - \d_{x-\hat\m,y} U^\dagger_{y\m} \right] \s_\m \,,
    \NON
    W_{xy} & = & 4\d_{xy} -\half \sum_\m \left[\d_{x+\hat\m,y} U_{x\m}
               + \d_{x-\hat\m,y} U^\dagger_{y\m} \right] \,, \nonumber
\end{eqnarray}
in which $\s_\m=(\vec\s,i)$, where $\s_k$ are the three Pauli
matrices, and each entry is a $2\times 2$ matrix.
The hermitian Wilson--Dirac operator
is $H(m_0)=\gamma_5D(m_0)$, with $\gamma_5={\rm diag}\;(1,1,-1,-1)$.
The theory is symmetric under the replacement $am_0\to -(8+am_0)$
because of the fact that for $am_0=-4$ the Wilson--Dirac operator
only contains nearest-neighbor couplings, thus allowing for a
$U(1)_\e$ symmetry \cite{SV}.

We define $H_0(m_0)$ as the operator $H(m_0)$ with all $U_{x\m}=1$.
This corresponds to the line $g_0=0$ in the phase diagram.
The spectrum of $H^2_0(m_0)$ covers a closed interval
$[(\l^0_{min}(m_0))^2,(\l^0_{max}(m_0))^2]$, with $0\le \l^0_{min}(m_0)
<\l^0_{max}(m_0)<\infty$.  $\l^0_{min}$ is determined by minimizing
\begin{equation}
  a^2 H_0^2(p;m_0) = \sum_\m \sin^2(a p_\m) + \left(
  \sum_\m(1-\cos{(ap_\m)})+am_0\right)^2
\label{free}
\end{equation}
over the Brillouin zone.  Keeping three components of the
momentum fixed, it is easy to see that $a^2 H_0^2(p;m_0)$
is linear in the cosine of the fourth component, and thus
minimized when this cosine equals $\pm 1$.  It follows that at
a minimum all four components of the momenta have to be equal
to $0$ or $\pi$, and thus that
$a^2 H_0^2(p;m_0)$ is the minimum over $k$ of $(2k+am_0)^2$, in which
$k$ is the number of momentum components equal to $\pi$.
We thus find that $\l^0_{min}={\rm min}\;|m_0-m_0^c|$, where
$am_0^c$ is one of the values $0$, $-2$, $-4$, $-6$, and $-8$.
We see that $\l^0_{min}$ is generically of order $1/a$, except
when $m_0$ is close to one of these critical points.  These
critical points correspond to the tips of the Aoki ``fingers"
on the line $g_0=0$ in Fig.~1.  Near these critical points the
theory yields 1, 4, 6, 4, and 1 light quarks per Wilson fermion,
respectively.  All eigenmodes of $H^2_0(m_0)$ are plane waves,
and thus extend over the whole volume.  Note that the spectrum
of $H_0(m_0)$ is contained in two disjunct intervals,
$[-\l^0_{max},-\l^0_{min}]$ and $[\l^0_{min},\l^0_{max}]$, separated
by a gap (if $\l^0_{min}\ne 0$).

Returning to the interacting theory, we consider for which
values of $m_0$ the operator $H(m_0)$, or equivalently $D(m_0)$, can have
zero eigenvalues.  Write $D(m_0)=A+iB$, with $A$ and $B$
hermitian, and consider an eigenmode $D\Psi=(A+iB)\Psi=\lambda\Psi$.
We then have that $\Psi^\dagger(A-iB)=\Psi^\dagger\lambda^*$,
and thus $2\Psi^\dagger A\Psi=(\lambda+\lambda^*)\Psi^\dagger\Psi$.
It follows that $\lambda$ can only vanish if $\Psi^\dagger A\Psi$
vanishes.  Since $A=-(W+am_0)$, this can only happen if
$-8\le am_0\le 0$, because the spectrum of $W$ is confined to the
interval $[0,8]$.  The super-critical region (\ie\ the region
where zero modes may exist) is thus the region
in which $-8\le am_0\le 0$, and we will restrict ourselves to that
region for the rest of this paper.

In the super-critical region ``exceptional" configurations may, and
do, occur.  Exceptional configurations are defined by the condition
that $D(m_0)$ has an eigenmode $\J_0$ with an exactly real
eigenvalue $\l_0$ \cite{SV,excep}.
For such configurations we have that $H(m_0+\l_0)
\J=\g_5 D(m_0+\l_0)\J=0$.  Hence, a configuration is exceptional
{\it iff} $H$ has an exact zero mode for some $m_0$.  BNN
configurations (see Sect.~4) are a special kind of exceptional
configurations.

We will be interested in a two-flavor theory constructed with
this Wilson--Dirac operator, with a fermion lagrangian
\begin{eqnarray}
\label{L}
\cl&=&\bj (D(m_0)-im_1\g_5\t_3)\j \\
&=&\bj'(H(m_0)-im_1\t_3)\j\,,\nonumber
\end{eqnarray}
where $\bj'=\bj\g_5$ is a field redefinition with jacobian one.
The (eight-component)
field $\j$ transforms in the fundamental representation of
isospin SU(2).  We added a symmetry-breaking term proportional to
$m_1$ pointing in the $\t_3$ direction, where $\t_k$
is another set of Pauli matrices acting in isospin space.
This symmetry-breaking term, which breaks both isospin and
parity, and thus has the quantum numbers of one of the pions,
allows us to probe the existence of an Aoki phase, in which
this pion field develops a vacuum expectation value.
We will assume the standard plaquette action for the gauge
field, unless otherwise noted.  Integrating over the fermion
fields yields the fermionic partition function
\begin{equation}
  Z_F = \prod_n (\l_n +im_1) (\l_n - im_1)  = \prod_n (\l_n^2 + m_1^2)\,,
\label{detF}
\end{equation}
where the product is over the eigenvalues $\lambda_n$ of
$H(m_0)$, of which there is a finite number on a finite-volume
lattice.  Note that $Z_F$ depends on the gauge-field configuration
through the eigenvalues $\lambda_n$.

The relevant Ward identity is obtained by performing a local
flavor transformation
\begin{equation}
\label{ISO}
\d_+\j(x)=i\a(x)\t_+\j(x)\,,\ \ \
\d_+\bj(x)=-i\a(x)\bj(x)\t_+\,,
\end{equation}
in which $\t_\pm=(\t_1\pm i\t_2)/2$.  With $\p_\pm(x)=
i\bj(x)\g_5\t_\pm\j(x)$ and $\p_3(x)=
i\bj(x)\g_5\t_3\j(x)$, we find for any operator $\co$ that
\begin{equation}
  \db_\m \svev{J^+_\m(x)\,\co(y)} + 2 m_1 \svev{\p_+(x)\,\co(y)}
    = {i\d_{xy}\over a^4} \svev{\d_+\co(y)} \,,
\label{gen}
\end{equation}
where the backward lattice derivative is defined by
$ \db_\m f(x) = ( f(x) - f(x-\hat\m))/a$
and the corresponding vector current is
\begin{equation}
J^+_\m(x) = \half \left(\bj(x) \t_+(\g_\m+1) U_\m(x)\j(x+\hat\m)
  + \bj(x+\hat\m) \t_+(\g_\m-1) U_\m^\dagger(x)\j(x) \right) \,.
\label{J}
\end{equation}
While the notation $\langle\dots\rangle$ indicates an integration
over both fermion and gauge fields, we note that the Ward identity
is also valid if we integrate over the
fermion fields only.  Taking $\co(y)=\p_-(y)$ and defining
\begin{eqnarray}
\label{Gammas}
\G(x,y)&=&\svev{\p_+(x)\,\p_-(y)} \,,\\
\G_\m(x,y)&=&\svev{J_\m^+(x)\,\p_-(y)} \,,\nonumber
\end{eqnarray}
we arrive at the Ward identity
\begin{equation}
\label{pp}
\db_\m\G_\m(x,y) + 2 m_1\G(x,y)={\d_{xy}\over a^4}
\svev{\p_3(y)}\,.
\end{equation}
Defining
\begin{equation}
\label{Gammap}
\tG(p) = {a^8\over V} \sum_{xy}
e^{ip(y-x)}\G(x,y) \,,
\end{equation}
and similarly $\tG_\m(p)$, the Fourier-transformed Ward identity is
\begin{equation}
  {1\over a} \sum_\m (1-e^{-iap_\m}) \tG_\m(p) + 2m_1 \tG(p)
  = \svev{\p_3} \,.
\label{ppft}
\end{equation}
The pionic condensate
\begin{equation}
\label{V}
\svev{\p_3} = (a^4/V) \sum_y \svev{\p_3(y)}
\end{equation}
can be calculated by first calculating the expectation value
of $\p_3(y)$ in a fixed gauge field (denoted by
$\langle\dots\rangle_\cu$) from $Z_F$,
\begin{equation}
\label{aoki}
a^4 \sum_y \svev{\p_3(y)}_\cu
    = Z_F^{-1} {\partial\over \partial m_1} Z_F
  =  2\, \sum_n {m_1\over \l_n^2  + m_1^2} \,,\nonumber
\end{equation}
and then averaging this over the gauge field, to obtain
\begin{equation}
   \svev{\p_3} = 2 \int d\l\, \rho(\l)  {m_1\over \l^2  + m_1^2} \,,
\label{pBC}
\end{equation}
where $\rho(\l)$ is the spectral density defined below.
In the limit $m_1\to 0$, we obtain
\begin{equation}
  \svev{\p_3} = 2\p \r(0) \,,
\label{BC}
\end{equation}
the Banks--Casher relation \cite{bankscasher} for the case at hand.

We define the spectral density from the cumulative density
\begin{equation}
  \cn(\l) = \svev{\sum_{\l_n\le \l} |\J_n(x)|^2} \,,
\label{cn}
\end{equation}
where $\J_n(x)$ is the eigenfunction associated with the eigenvalue $\l_n$,
\begin{equation}
\label{Heigen}
H(m_0)\J_n=\l_n\J_n\,,
\end{equation}
and with normalization $a^4\sum_x|\J_n(x)|^2=1$.  Because of the gauge-field
average in Eq.~(\ref{cn}) the right-hand side is actually
independent of $x$.  The spectral density $\r(\l)$ is
\begin{equation}
  \r(\l)=d\cn/d\l= V^{-1} \svev{\sum_n \d(\l-\l_n)} \,.
\label{rho}
\end{equation}
All these results are valid in finite-volume QCD, in both the
quenched and unquenched cases.  They remain true if the
infinite-volume limit is taken, but of course this limit may
or may not commute with the limit $m_1\to 0$.  It is well known
that in unquenched QCD these limits do not commute in a phase
with SSB; the order parameter $\svev{\p_3}$ in the $m_1=0$
theory vanishes in finite volume, but does not vanish if
the thermodynamic limit is taken before the limit $m_1\to 0$.
We will return to this point in Sect.~5.

For any given $m_0$ we will denote by $\l^2_{min}$ ($\l^2_{max}$)
the minimum (maximum) eigenvalue of $H^2(m_0)$ over the
gauge-field space. These values are finite,
because of the fact that $H(m_0)$ connects a given site only to a finite
number of neighboring sites, and because the gauge group is compact;
hence $H(m_0)$ is uniformly bounded.
The cumulative density $\cn(\l)$, defined by the spectrum of $H(m_0)$,
is monotonously non-decreasing on the interval $[-\l_{max},\l_{max}]$.
Using the results of Ref.~\cite{BNN}, we will argue below that
$\l_{min}=0$ for any $0>am_0>-8$ and that, in the quenched theory,
$\r(\l)>0$ for any $-\l_{max}<\l<\l_{max}$.
We will first consider a finite volume (Sect.~3) and then
the infinite-volume limit (Sect.~4).
Outside the super-critical region, \ie\ for $am_0>0$ or $am_0<-8$,
one has $\l_{min}>0$, and $\r(\l)=0$ for $-\l_{min}<\l<\l_{min}$.

\vspace{5ex}
\noindent {\large\bf 3.~Spectral representation of the pion
two-point function}
\secteq{3}
\vspace{3ex}

We are now ready to establish one of the key results.
Using the spectral representation for the quark propagators,
it is straightforward to derive that, in a finite volume,
\begin{equation}
  \G(x,y) = \int \cd\cu \cb(\cu) \sum_{kn}
      \J_n^\dagger(x) \J_k(x) {1\over \l_k + im_1}
      \J_k^\dagger(y) \J_n(y) {1\over \l_n - im_1} \,.
\label{dblsum}
\end{equation}
The two propagators on the right-hand side correspond to the ``up'' and
the ``down'' quarks respectively. Each sum runs over the eigenstates of
the (single-flavor) hermitian Wilson--Dirac operator $H(m_0)$.
In full QCD the Boltzmann weight
is $\cb= Z_F \exp(-S_g)$, where $S_g$ is the gauge action,
and the fermionic partition function $Z_F$
is given in Eq.~(\ref{detF}). In quenched QCD, $\cb=\exp(-S_g)$.

For a generic gauge field, the eigenvalue spectrum of $H(m_0)$
is non-degenerate: if $\l_n=\l_k$ then $\J_n(x)=\J_k(x)$.
Gauge field configurations for which this is not true form
a subset of the configuration space with measure zero,
because either $H(m_0)$ would have to have a symmetry in this
fixed gauge-field background, or there is an accidental
degeneracy.  Both types of degeneracy disappear under a
generic small deformation of the gauge field.

In a finite volume on the lattice, the sums over $n$ and
$k$ in Eq.~(\ref{dblsum}) are finite, and the integration over
gauge fields is absolutely convergent because we integrate
over a compact group.  This implies that we may interchange
the sums and the integral, and consider separately the contribution for
each $n$ and $k$ to $\G(x,y)$, which is
\begin{equation}
  \int \cd\cu \cb(\cu)
      \J_n^\dagger(x) \J_k(x) {1\over \l_k + im_1}
      \J_k^\dagger(y) \J_n(y) {1\over \l_n - im_1} \,.
\label{term}
\end{equation}
For $n\ne k$ (and thus generically $\l_n\ne\l_k$),
expression~(\ref{term}) is finite, even in the limit $m_1\to 0$.
This is because the denominators $\l_n+im_1$ and $\l_k-im_1$
become singular on different subspaces (with codimension one)
of the gauge-field space,
and the $im_1$ terms in the denominators provide an
$i\epsilon$ prescription for how to integrate around the
poles at $\l_n=0$ or $\l_k=0$.  Therefore, in the limit $m_1\to 0$ only terms
with $n=k$ can make a non-zero contribution to the product
$m_1 \G(x,y)\;$:
\begin{equation}
  \lim_{m_1 \to 0} m_1 \G(x,y) =
    \lim_{m_1 \to 0} \int \cd\cu \cb(\cu) \sum_n
      |\J_n(x)|^2 |\J_n(y)|^2 {m_1 \over \l_n^2 + m_1^2} \,.
\label{limm1}
\end{equation}
Of course, the fact that we keep the volume finite is a
key element of this argument.  For the Fourier transform of
$\G(x,y)$, Eq.~(\ref{Gammap}), this result translates into
\begin{eqnarray}
\label{cf}
  \lim_{m_1 \to 0} m_1 \tG(p) &=&
    \lim_{m_1 \to 0} {m_1\over V} \int \cd\cu \cb(\cu)
      \sum_n { |\ch_n(p)|^2   \over \l_n^2 + m_1^2}\,,\\
      \ch_n(p) &=& a^4\sum_x |\J_n(x)|^2 e^{-ipx} \,.\nonumber
\end{eqnarray}
Eq.~(\ref{limm1}) has a dramatic consequence for quenched
QCD.  We may introduce the density correlation function
\begin{equation}
  \car(x,y;\l) = \int \cd\cu \cb(\cu) \sum_n
    |\J_n(x)|^2 |\J_n(y)|^2 \d(\l-\l_n) \,,
\label{Rdef}
\end{equation}
with which Eq.~(\ref{limm1}) can be written as
\begin{equation}
  \lim_{m_1 \to 0} m_1 \G(x,y) =
  \lim_{m_1 \to 0}  \int d\l\, \car(x,y;\l) {m_1 \over \l^2 + m_1^2}
  = \p \car(x,y;0) \,.
\label{limRdef}
\end{equation}
It is clear that $\car(x,y;\l)$ is independent of $m_1$ in the
quenched theory because $\cb(\cu)$ is.
Also, if we choose $m_0$ anywhere inside the super-critical region,
the existence of BNN-like zero modes\footnote{
  For a more detailed account of BNN's work, see Sect.~4.
}
implies that there is no spectral gap around $\l=0$,
and, in particular, that exact zero modes
occur on a non-empty subspace of codimension one.
As a result, $\car(x,y;\l)>0$ for any $-\l_{max}<\l<\l_{max}$, because,
apart from the delta function, the integrand in Eq.~(\ref{Rdef}) is
strictly positive.
Hence, we find a $1/m_1$ divergence in $\G(x,y)$,
\begin{equation}
  \G(x,y) = {\p \car(x,y;0)\over m_1} \; + \; O(1) \,.
\label{Rm1}
\end{equation}
Moreover, if $\car(x,y;0)>0$, also $\r(0)>0$ and, from Eq.~(\ref{BC}),
$\svev{\p_3}>0$ for the same reasons,
and we find that we can have SSB in a {\it finite} volume in the
quenched theory.\footnote{For the unquenched theory, see Sect.~5.}
Eq.~(\ref{Rm1}) shows that the quenched theory in the
super-critical region
is singular in the limit $m_1\to 0$, and it provides an alternative
mechanism for saturating the Ward identity Eq.~(\ref{pp}).  With this
alternative mechanism, no Goldstone excitations need to appear
in the quenched theory, even in the presence of SSB.

The fact that these results are so simple is related to the fact
that, in the finite-volume
quenched theory, the existence of both the condensate
and the $1/m_1$ divergence in the pion two-point function are
kinematical effects, which depend only on the super-criticality
of $m_0$ and the strict positivity of the ($m_1$-independent)
quenched measure.  The magnitude of $\svev{\p_3}>0$ and
$ \car(x,y;0)$ depend on the dynamics, and, as it will turn
out, on whether we are inside or outside the Aoki phase.

To conclude this section, we note that one
can repeat the entire analysis assuming a different
SU(2)-orientation of the flavor- and parity-symmetry breaking term
in the lagrangian~(\ref{L}).  The orientations of both the pionic
condensate and the divergent term in the pion two-point function
follow the orientation of this symmetry breaking term, just
as in the case of conventional SSB.
In the next section, we will address the dynamics of the quenched
theory, and in the section after that the differences with the
unquenched theory.

\vspace{5ex}
\noindent {\large\bf 4.~Localization and spontaneous symmetry
breaking in the quenched theory}
\secteq{4}
\vspace{3ex}

In a given finite volume, all eigenmodes of $H(m_0)$ are localized
for the trivial reason that their support is compact.  In
particular, all near-zero modes contribute to the $1/m_1$ divergence
we found in the previous section.  (Here we define a near-zero
eigenvalue as an eigenvalue with an absolute value of order
$m_1$, for a given (small) $m_1$.)  However, as we will now
demonstrate, quantitatively the finite-volume divergence
comes from exponentially localized near-zero modes.  We postpone
the discussion of what happens in the infinite-volume limit
to the second part of this section.

\vspace{5ex}
\noindent {\large\it 4.1~Localized modes and the divergence of
the pion two-point function}
\vspace{3ex}

On a finite lattice of spacing $a$ and (large) linear size $L$
(in all four directions), a normalized eigenstate
$\J_n(x)$ is exponentially localized, provided there exist a
positive constant $c_1=O(1)$, a lattice site $x_n^0$ (which depends
on  $\J_n$), and some $l$ with $a \le l \ll L$ such that
\begin{equation}
  | \J_n(x) |^2 \le \,
    {c_1\over l^4}\, \exp\left(- {|x-x_n^0| \over l}  \right) \,.
\label{lclz}
\end{equation}
The {\it localization range}  (or localization length)
$l_n$ is the minimal value of $l$
for which the bound~(\ref{lclz}) holds.  For definiteness,
the distance $|x-y|^2$ is
defined by the minimum of $\sum_\m |x_\m-y_\m+n_\m L|^2$ over all
integers $n_\m$.
Our intention is to establish a lower bound on $\tG(p)$ in
terms of the density of localized near-zero modes and, for this purpose,
we take $c_1$ to be independent of $l$ and $L$.
Also, the restriction to $l_n \le l \ll L$ means that we consider only
localized modes whose support (of roughly order $l_n^4$), and size
$|\J_n|^2 \sim 1/l_n^4$ inside this support,
are basically independent of the volume.  This definition may
exclude some modes which would be exponentially localized
according to some other reasonable definition,
but the class we are considering here will be sufficient for our purpose.

We begin with establishing bounds on $\ch_n(p)$ in Eq.~(\ref{cf}):
\begin{equation}
  1 - c_2\, l_n^2 p^2 \le |\ch_n(p)|^2 \le 1 \,.
\label{bound}
\end{equation}
Here $c_2 \propto c_1$ is another numerical constant which we
will define below.  The upper bound is trivial, while the lower
bound just expresses the fact that one cannot resolve the
structure of a localized mode with localization range $l_n$
using momenta $p\ll 1/l_n$.  The lower bound is established
by noting that (with $x_n^0$ the ``center" of the localized
eigenstate of Eq.~(\ref{lclz}))
\begin{eqnarray}
\label{ineq}
  0 \le 1- \Re (e^{ip x_n^0} \, \ch_n(p))
    & = & 2 a^4 \sum_x \sin^2(p(x-x_n^0)/2)\, |\J_n(x)|^2 \\
    & \le & {c_1\over 2} {a^4\over l_n^4}
     \sum_x  p^2 \, (x-x_n^0)^2\, e^{- \, |x-x_n^0|\, / \, l_n}
    \NON
    &=& {\cc(l_n)\over 2}\, l_n^2\, p^2
   \le  {c_2\over 2}\, l_n^2\, p^2 \,.
\nonumber
\end{eqnarray}
The inequalities~(\ref{lclz}) and $|\sin(\a)|\le \a$ were used.
The dimensionless quantity
\begin{equation}
\label{cl}
  \cc(l) = c_1 {a^4\over l^4} \sum_x z^2 \, e^{- |z|} \,
          \Big|_{z_\m =x_\m/l}
\end{equation}
is a continuous function of $l$ and finite in the limit $l\to\infty$.
The fact that $\cc(l)$ is finite for $l\to\infty$
is a consequence of choosing the bound on
$|\J_n(x)|^2$ with the prefactor $c_1/l^4$, with $c_1$ independent
of $l$.  This is important for deriving a non-trivial lower bound
on $\tG(p)$.
It thus follows that
\begin{equation}
  c_2 \equiv {\rm max}_{\,l\ge a} \{ \cc(l) \}
\label{maxc}
\end{equation}
is a finite numerical constant of the same order as $c_1$.
Since $|\Re (e^{ip x_n^0} \, \ch_n(p))|\le|\ch_n(p)|$ the
inequality~(\ref{bound}) follows.

We now use this bound to derive a lower bound on $\tG(p)$ in
Eq.~(\ref{cf}), for $m_1\to 0$. Choosing some fixed $l$, we split the
eigenstates of $H(m_0)$ into two classes:
those which are exponentially localized according to
our definition of Eq.~(\ref{lclz}) with localization range $l_n\le l$,
and the rest.  Keeping only the former,
we obtain
\begin{equation}
  m_1\tG(p)
    \ge {m_1\over V} \int \cd\cu \cb(\cu)
    \sum_{l_n \le l} { 1 - c_2\, l_n^2\, p^2 \over \l_n^2 + m_1^2}
     \; + \; O(m_1) \,.
\label{split}
\end{equation}
Since in this sum we only keep terms for which
$l_n\le l$, this may be expressed more simply as
\begin{equation}
  m_1\tG(p) \ge \p\r_l(0)\, \left(1 - c_2\, l^2 p^2 \right) +
  O(m_1) \,.
\label{div}
\end{equation}
Here $\r_l(\l)$ is the density of exponentially localized
eigenstates (according to our definition~(\ref{lclz})) with
eigenvalue $\l$ and localization range less than or equal to $l$.

This result is of key importance.  It establishes that if
there is a non-zero density of near-zero modes which are exponentially
localized within a certain range $l$, the pion two-point function
diverges as $m_1\to 0$ in any finite volume, for all momenta
$p$ below (roughly) the inverse localization range.  As we will
argue in Sect~4.2 below, the analysis of Ref.~\cite{BNN}
allows us to establish the same result in the infinite-volume limit,
provided the mobility edge does not vanish
(and $m_0$ is in the super-critical region).  This provides
an alternative mechanism for saturating the Ward identity,
Eq.~(\ref{ppft}), in the presence of a non-vanishing pion condensate
$\svev{\p_3}$.  It also shows that the
quenched theory in infinite volume is
not well defined for $m_1=0$, at least as long as the mobility
edge does not vanish.

A comment is in order on the range of momenta for which our
result is valid, because of the fact that in a finite volume
the momenta are discrete, and thus cannot be chosen arbitrarily small.
First, we observe that for a
localized near-zero mode, the scale of the typical
localization range $l$ is
set by the inverse mobility edge, $\l_c^{-1}$.
(This will be explained in more detail in the following subsection.)
As long as we are
not close to the Aoki phase, this is of order one in lattice
units.  With the smallest lattice momenta being of order
$1/L$, we may choose the volume large enough that
$c_2(l/L)^2\ll 1$, so that non-zero momenta indeed exist for
which the right-hand side of Eq.~(\ref{div}) is positive.

\vspace{5ex}
\noindent {\large\it 4.2~Spontaneous symmetry breaking without
Goldstone excitations}
\vspace{3ex}

We are now at a point where we wish to collect all the available results,
and use them to construct a conjecture for the quenched phase diagram.
The goal is to combine the numerical \cite{scri,jansenetal}
as well as the analytical \cite{BNN} evidence that basically
everywhere in the super-critical region the density of near-zero
modes does not vanish (already in finite volume) with the
inequality for the pion two-point function, Eq.~(\ref{div}).  In a
sense, we will be proposing the simplest possible picture of what
determines the phase diagram, while requiring consistency with
the Ward identity~(\ref{ppft}).  While we already summarized
our conjecture for the phase diagram in the introduction,
we present a more detailed discussion in this subsection.

A key element is the observation made by BNN in Ref.~\cite{BNN}, that
in the super-critical region there will be a
density of near-zero modes.  We will
therefore start with elaborating on this observation.  As
already alluded to in the introduction, it will
lead us to borrow the concept of the ``mobility edge" from
condensed-matter physics as the crucial ingredient in
determining the quenched phase diagram.

Following BNN, we begin with an infinite-volume configuration
containing a single dislocation. We pick
a lattice gauge field which is equal to the classical vacuum
($U_{x\m}=I$)
everywhere, except for the links of an $n^4$ hypercube
containing the origin. These links can take any value.
The square of the Wilson--Dirac operator can be written as
\begin{equation}
  H^2(m_0) = H_0^2(m_0) + V\,,
\label{H2}
\end{equation}
where $H_0(m_0)$ is the free operator.
The potential $V$ is supported on an $(n+2)^4$ hypercube.
The ``hamiltonian" $H^2(m_0)$ is a discretized,
semi-positive Schr\"o\-dinger operator with a finite-range potential,
with a continuum threshold at $(\l^0_{min})^2$ (\cf\ Sect.~2).
All eigenstates with $(\l^0_{min})^2 < E < (\l^0_{max})^2$ are
scattering states,
while those with $0\le E < (\l^0_{min})^2$, if they exist, are bound
states
of $H^2(m_0)$ with  ``binding energy''
\begin{equation}
  -E_b \equiv E-(\l^0_{min})^2 < 0 \,.
\label{bind}
\end{equation}
Of course, bound-state wave functions decrease exponentially
away from the hypercube,
with a decay rate determined by $E_b^{-1}$.  In particular,
a zero mode has $E=0$, and thus a binding
energy $E_b=(\l^0_{min})^2$.

For single-dislocation configurations, BNN proved that the problem
of finding a zero mode in large (or infinite) volume
can be reduced to a corresponding eigenvalue problem
on the small hypercube.
Solving the latter problem numerically BNN then found that,
for $-1 > a m_0 > -7$ (roughly),
the links of a hypercube as small as $2^4$ can always
be chosen such that $H(m_0)$ has an eigenvalue equal
(or numerically extremely close) to zero.
This means that, for that range of $m_0$, zero modes
exist for any volume down to $2^4$.
Moreover, these zero modes are exponentially localized
(as long as $m_0$ is not equal to one of its critical values).
We note that BNN configurations are examples of
exceptional configurations.\footnote{Even though we expect BNN
configurations to amount to only a tiny subset of all exceptional
configurations.}

It is not implausible that, by allowing for larger dislocations,
exponentially-localized zero modes would be found throughout
virtually the entire super-critical interval $0 > a m_0 > -8$.
One does expect that, to get a zero eigenmode closer to a
boundary point of the super-critical interval, the gauge field
would have to differ from the classical vacuum over bigger hypercubes.
In this paper, we will assume that this is the case, \ie\ that
for any $am_0$ strictly inside the interval $(-8,0)$, BNN-like
zero modes will be found.

Before continuing, let us comment on the nature of typical zero modes,
and the configurations supporting them.
An exact zero mode of $H^2(m_0)$ is also a zero mode of $H(m_0)$.
As one varies the lattice gauge field,
a zero eigenvalue generically corresponds to a zero-level crossing,
because with Wilson fermions there is no chiral symmetry that might
protect a zero eigenvalue from moving away from zero.
In other words, if $\cu(\t)\equiv\{U_{x\m}(\t)\}$
is a family of gauge-field configurations
that depend smoothly on $\t$ , and if there is an eigenvalue
$\l_n$ such that $\l_n(\cu(0))=0$ then, typically,
$\l_n(\cu(\t))$ changes sign at $\t=0$.
This is true for (almost) all exceptional configurations
which support exact zero modes. It implies that any finite-volume
configuration supporting an (exponentially localized) exact zero mode
has some open neighborhood, in the gauge-field space,
with a spectrum of near-zero modes.
Also, in a finite volume,
the subspace of configurations that support an exact zero mode
has codimension one.

Our next step is to consider configurations containing
a very small density of BNN-like dislocations.
The dislocations are small, surrounded by the
classical vacuum, and (still, on average) far apart.
This is a ``controlled"
form of {\it disorder}. The origin of disorder is two-fold:
the positions of the dislocations
are chosen at random (one can speak about a dilute gas of dislocations);
also the links that define each dislocation are chosen at random.
A unique limiting value $\lim_{|x|\to\infty} V(x)$ of the potential
in Eq.~(\ref{H2}) no longer exists, because dislocations may be
found arbitrarily far from the origin.

As long as the gas of dislocations is dilute, we can still
identify all dislocations separately from the surrounding vacuum,
simply by inspection.  So, let us focus on one dislocation,
and assume that in isolation (\ie\ with no other dislocations
anywhere), it has an exact zero mode.  Since the other dislocations
are far apart, they will have a negligible effect on this
zero mode.  This is easily seen by invoking a variational
argument.  Using the eigenfunction of the zero mode
of the isolated dislocation as a trial wave function, the expectation
value of $H^2(m_0)$ on this trial state
will receive contributions only from the other distant
dislocations.  If the mean distance between dislocations is $R$,
the expectation value of $H^2(m_0)$
will be of order $e^{-2\g}$ where $\g\sim \l^0_{min} R$.
Therefore $H^2(m_0)$ must have
an eigenstate with an exponentially small eigenvalue
$E \sim e^{-2\g}$
which closely resembles the original zero mode.
We may now vary the gauge field at the dislocation such that
the eigenvalue under consideration varies as well.  As long
as $R$ is large enough (compared to $1/\l^0_{min}$), $e^{-2\g}$
is small enough that this gauge-field variation will vary
the eigenvalue over an interval that includes zero.

The situation changes qualitatively if the density of
dislocations is large, and also will be different
for configurations generated in a typical Monte-Carlo
simulation.  In order to describe this situation, it will
be useful to introduce the {\it mobility edge hypothesis} \cite{lcl}.
The mobility edge is well-defined only
in infinite volume, and we will thus assume the volume to
be infinite for the following discussion.  The hypothesis states
that, when disorder
is introduced in an ordered system, the conduction-band
structure of the ordered system is replaced by a number
of alternating energy intervals, each containing either
(exponentially) localized or extended eigenstates.  In
particular, no energy exists for which there are both
localized and extended eigenstates.\footnote{
  We are not aware of any proof of this---widely used---assertion.
}
The energy separating
an interval with localized modes from one with extended
modes is referred to as a mobility edge.

Intuitively, a mobility edge arises as follows.
Let us first return to a situation with ``controlled'' disorder
consisting of a dilute gas of BNN dislocations.
As already mentioned, for a very small density of dislocations,
all bound states with $E_b=O(1/a^2)$ (see Eq.~(\ref{bind}))
will be exponentially localized near a single dislocation.
As the density of dislocations is increased there is
an increased probability that an ``electron,'' situated initially in a
bound state of a given dislocation,
will tunnel into a near-by dislocation.
When the probability to tunnel a certain distance
times the average number of available dislocations within this distance
has become $O(1)$ for some given $E$, the electron will be able to
travel infinitely far by tunneling from dislocation to dislocation,
regardless of the
details of the dislocation that produced the original bound state.
In other words, the eigenstate with eigenvalue $E$ becomes extended.

For a single dislocation, any eigenstate
with $E<(\l^0_{min})^2$ decays exponentially.
Increasing $E$ means a slower decay rate
and, hence, an increased tunneling probability.
This means that eigenstates with higher $E$ become extended first
when the density of dislocations is increased.
If $\l_c^2$ is the mobility edge, the range $0\le E \le \l_c^2$
will consist of exponentially localized eigenstates,
while the range $E \ge \l_c^2$ consists of extended ones.
For a very small density of dislocations the mobility edge
will be close to $(\l^0_{min})^2$, while for larger densities it will
move further away from this value.
If, for instance, we allow only dislocations generating an
attractive potential
in Eq.~(\ref{H2}), the mobility edge will clearly be lower than
$(\l^0_{min})^2$.

We note that there may exist more than one mobility edge at a
given location in the phase diagram.  For instance, if the
hamiltonian is bounded from above, one expects another mobility
edge to occur, {\it above} which other localized eigenstates
exist.  In our case, we expect a second mobility edge like
this, because the eigenvalue spectrum of the free hamiltonian,
$H_0^2(m_0)$ in Eq.~(\ref{free}), also has a maximum eigenvalue
$(\l^0_{max})^2$.  In this paper, we will be concerned with the
lowest-lying mobility edge, and in particular, the question of
whether it vanishes or not.

Now consider the hypothetical situation that,
for a certain gauge-field configuration, both
extended and localized eigenstates of the hamiltonian $H^2(m_0)$
occur at the same energy $E$.  If we then calculate the eigenstates
on a typical small fluctuation of this gauge field, the extended
and localized states will mix, and all new eigenstates will be
extended.  The only possible exception is when all eigenstates of the
original gauge field were localized.  A ``typical" configuration thus
has either only extended or only localized eigenstates at a given
$E$.  Since typical configurations (in this sense) determine the
properties of an ensemble, it follows that the value of a
mobility edge separating localized and extended eigenstates is
associated with an equilibrium ensemble, and thus with a point
$(m_0,g_0)$ in the phase diagram.  Another way of arriving at the
same conclusion is to note that
in the thermodynamic limit, one equilibrium
configuration suffices to determine the equilibrium properties
of the theory, and thus is typical.
For our purposes, it will be useful to characterize each
point $(am_0,g_0)$ in the phase diagram
by the value of the lowest mobility edge of $H^2(m_0)$.
(In the rest of this paper, we will refer to this mobility edge
as ``the" mobility edge.)  The mobility-edge hypothesis
will be at the heart of our conjecture for the phase diagram of
quenched QCD.

Before we get to this conjecture, let us briefly comment on
the average localization range as a function of $E$.  For energies
at which
only localized states exist, one expects to be able to define
an average localization range $\bl=\bl(E)$.  This average
localization range should then grow if the energy $E$ approaches
the mobility edge, and diverge at and beyond the edge, where
only extended eigenstates exist.

Based on BNN's results we expect
that anywhere inside the super-critical region
there will be a non-zero density of near-zero modes.
If the mobility edge is zero this expectation is fulfilled by assumption.
If the mobility edge is larger than zero, then,
for any $0>am_0>-8$ and $g_0>0$,
there is a finite probability per unit volume that any given
Monte-Carlo configuration
will contain a ball with (a large) radius $R$, inside which the
configuration
will resemble the classical vacuum, apart from a single BNN dislocation
well inside that ball.  Using the same variational argument as in
the dilute-gas case, the exact BNN zero mode associated with the
dislocation implies the existence of an eigenstate with
eigenvalue $e^{-2\g} \ll \l_c^2$, well below the mobility edge.
This eigenstate decays exponentially both inside
and outside of the classical-vacuum ball.  By invoking small
deformations of this configuration, the existence of a density of
exponentially localized near-zero modes follows.\footnote{
  A zero-level crossing in the spectrum of $H(m_0)$
  changes the index of the overlap operator.
  This index was advocated as a definition of the topological charge
  on the lattice \cite{oovlp,ovlp}. However, standard arguments imply
  that the topological charge density scales like $1/\sqrt{V}$
  in large volume, and, thus, vanishes for $V\to\infty$
  (see \eg\ Ref.~\cite{LS}). Therefore,
  topological considerations do not explain a non-zero
  density of near-zero modes in the infinite volume limit.
  This situation reflects the inherent ambiguity in trying to identify
  small dislocations as instantons or as anti-instantons.
}

We now formulate our main conjecture about the phase diagram
of quenched QCD with two Wilson fermions.  Near the line $g_0=0$,
we expect the mobility edge to be very close to $(\l^0_{min})^2$.
In particular, with $(\l^0_{min})^2$, it will vanish at the critical
``end" points $g_0=0$, $am_0=0$, $-2$, $-4$, $-6$, $-8$.
As $g_0$ is increased and the typical gauge-field configuration
in an equilibrium ensemble
becomes less smooth, the mobility edge will move away from
$(\l^0_{min})^2$, and, in parts of the phase diagram, it may vanish.
Since it vanishes at the critical end points on the $g_0=0$ axis,
it is reasonable to expect that the mobility edge will also vanish
in the vicinity of those points for $g_0>0$, thus opening up
the finger-like structure of region B in Fig.~1.  In a region
where the mobility edge vanishes, the density of near-zero modes
will be due to extended near-zero modes, and one expects
long-range behavior, in particular Goldstone excitations.
We thus conjecture that, in quenched QCD,
the Aoki phase---which is defined as the phase
in which Goldstone excitations exist---coincides
with the region of the phase diagram where the
mobility edge vanishes.  The near-zero mode density, and thus
the value of $\svev{\p_3}$, is due to extended eigenstates.
Outside the Aoki phase (but inside the super-critical region),
the mobility edge is larger than zero, and the density is due
to exponentially localized near-zero modes.  Because of this
non-vanishing density, the mechanism derived in the previous
subsection will kick in, and the pion two-point function will
diverge for $m_1\to 0$ for small-enough momenta.

Extended states should not contribute to the $1/m_1$ divergence
in the infinite-volume limit.
The number of extended states grows like the volume $V$, whereas
the contribution to Eq.~(\ref{limm1}) of each extended state drops
like $1/V^2$.  This yields a contribution of order $1/V$,
which vanishes for $V\to\infty$.  Therefore, we expect
no $1/m_1$ divergence of the pion two-point function inside
the Aoki phase; the $1/m_1$ divergence characterizes
region C (and the super-critical part of region A) of the
phase diagram.  It follows that,
while $\svev{\p_3}$ is not a useful order parameter
for detecting the Aoki phase, the residue of a $1/m_1$ divergence in
$\tG(p)$ (for small enough momentum) is a useful order
parameter in this sense.  There is even a {\it local} order
parameter associated with this divergence:
\begin{equation}
  \x(x) \equiv \lim_{m_1\to 0} m_1 \G(x,x)
      = \lim_{m_1\to 0} m_1 \svev{\p_+(x)\,\p_-(x)} \,.
\label{xi}
\end{equation}
This follows from $\x(x)=(1/V)\sum_p\tG(p)$, the bound~(\ref{div}),
and the positivity of $\tG(p)$ for all $p$.

We believe that essentially the same conjecture holds for
{\it unquenched} QCD. The only difference is that in the unquenched
case, the spectral density $\r(0)$
(and thus the order parameter) vanishes outside
the Aoki phase, because of the suppression of the localized
near-zero modes by the fermion determinant (\cf\ Sect.~5).
Therefore, also the $1/m_1$ divergence in the pion two-point
function will not occur in the unquenched case.  These
phenomena are quenched artifacts, but the existence and role
of the mobility edge are not.
To summarize, the qualitative features of the phase diagram
in Fig.~1 are valid in both quenched and unquenched QCD.
The mobility edge $\l_c$ is zero in region B,
and non-zero outside of this region,
where exponentially localized modes with $\l^2<\l_c^2$ exist.
In regions C and in the super-critical part of regions A
the spectrum of localized modes extends down to zero,
with the only difference that $\r(\l)>0$
for $\l^2\ge 0$ in the quenched theory,
and for $\l^2>0$ in the unquenched theory.

Clearly, our picture of the phase diagram proposed here is
a conjecture, and we have no proof that it is correct.
We emphasize that it appears to be the simplest possible
way one can understand the phase diagram, given the available
analytical and numerical evidence.  We will end this section
with a more quantitative, but still heuristic argument as
to why Goldstone excitations {\it only} occur in the
phase with vanishing mobility edge.

We will assume that outside the Aoki phase, the near-zero-%
mode density $\r(0)$ is entirely due to localized zero modes
captured by our bound, Eq.~(\ref{lclz}).  Moreover, we will assume
that, for practical purposes, a maximum localization length
$l_{max}$ exists for the near-zero modes.  Hence\footnote{
  Recall that $\r_l(\l)$ is the density of localized eigenstates with
  localization range less than $l$.
}
$\r_{l_{max}}(0)=\r(0)$, and,
using both the lower and upper bounds in Eq.~(\ref{bound}),
\begin{equation}
  \p\r(0) (1 - c_2\,l_{max}^2 p^2)
    \le \lim_{m_1\to 0} m_1 \tG(p) \le  \p\r(0) \,.
\label{singlr}
\end{equation}
Writing $\tG_\m(p)=ip_\m\S(p^2)$, it then follows from the
Ward identity, Eq.~(\ref{ppft}) that
\begin{equation}
\S(0)=2\p\r(0)\, O(l_{max}^2) \,,
\label{nopole}
\end{equation}
leading to the conclusion that $\tG_\m(p)$ does not have a massless
pole.

The most drastic assumption we make here is that there is a
finite $l_{max}$.  This is rather unlikely, even if it is
reasonable to expect that the probability to find a near-zero
mode with a very large localization range at some point well
outside the Aoki phase is very small.  In order to argue that
also in this case no Goldstone poles occur outside the Aoki
phase, clearly one needs a better estimate of $|\ch_n(p)|$
than that provided by
the bounds given in Eq.~(\ref{bound}).  While it is a very
hard problem to come up with better estimate, we will use
an {\it ansatz} for a better estimate to show how an improved
argument might work.  Let us assume that $|\ch_n(p)|$ can be
estimated by
\begin{equation}
  |\ch_n(p)| \,\sim\, {1\over 1 + p^2 l_n^2} \;, \qquad pa \ll 1 \,.
\label{Hp}
\end{equation}
This is not unreasonable for momenta $p\ll 1/a$.  For very small
momenta, one expects only the long-distance features of the
near-zero modes, \ie\ the exponential tail in Eq.~(\ref{lclz}),
to determine the Fourier transform of $|\J_n(x)|^2$, and thus
$|\ch_n(p)|$ would have to look something like this.
The precise form is not important; the crucial properties of this
{\it ansatz} are that it is bounded for arbitrary $l_n$, and
that it is consistent with the fact that $|\ch_n(0)|=1$, as
follows from the fact that $\J_n(x)$ is normalized to one.

If $\r(0)$ is completely due to localized near-zero modes,
we may write $\r(0)=\int_0^\infty dl\,\r'_l(0)$ with
$\r'_l(\l)=d\r_l(\l)/dl$, and Eq.~(\ref{cf}) as
\begin{eqnarray}
\label{intll}
\lim_{m_1\to 0} 2m_1\tG(p)&\sim &
{2\p\over V}\int_0^\infty dl \int \cd\cu \cb(\cu)
\sum_n\d(\l_n)\d(l-l_n)\,{1\over (1 + p^2 l^2)^2} \\
&=&2\p \int_0^\infty dl \,\r'_{l}(0)
\,{1\over (1 + p^2 l^2)^2} \,.
\nonumber
\end{eqnarray}
Combining this with Eqs.~(\ref{ppft}) and ~(\ref{BC}) we obtain
the estimate
\begin{equation}
  \S(p^2)  \sim
  2\p \int_0^\infty dl\,
  \r'_{l}(0)\, {2l^2 +p^2 l^4 \,\over (1 + p^2 l^2)^2} \;,
\label{Spx}
\end{equation}
showing again that $\tG_\m(p)$ has no massless pole.  We note that
the integral over $l$ should be convergent because
$d\r_l(0)/dl$ should be strongly suppressed for large $l$.

This completes our discussion of the quenched phase diagram.
In the remainder of this paper, we will expand on the differences
between the quenched and unquenched cases, and explore the
consequences for domain-wall and overlap fermions.

\vspace{5ex}
\noindent {\large\bf 5.~Unquenched QCD}
\secteq{5}
\vspace{3ex}

In this section we only review well-known facts about unquenched
QCD.  The only reason we include this brief review is to point
out that none of the surprising results we obtained in the
previous section for quenched QCD are in conflict with the
standard lore in the unquenched theory.

\vspace{5ex}
\noindent {\large\it 5.1~Vanishing of the condensate
in finite volume and absence of localized near-zero modes}
\vspace{3ex}

We start with a review of the proof that the finite-volume
condensate is zero in unquenched QCD.  Summing the Ward
identity~(\ref{pp}) over space-time, the total-derivative term
drops out, and we obtain
\begin{equation}
  2 m_1 a^4 \sum_x \svev{\p_+(x)\,\p_-(y)}
    =  \svev{\p_3} \,.
\label{pp0}
\end{equation}
In a finite volume, the correlation function of the product of any
(finite) number of fermion (and link) operators is bounded.
The left-hand side of Eq.~(\ref{pp0}) is therefore the product of $m_1$ with
a bounded function, hence it vanishes for $m_1 \to 0$.
It follows that the condensate vanishes.  This is the familiar result
that there is no spontaneous symmetry breaking in a finite volume.

In order to prove boundedness, consider first the unnormalized
expectation value of any observable made of a product of (link and)
fermion fields.  Because of the Berezin integration rules for
Grassmann variables, the Wick contraction of the fermion fields
leads to a function which is analytic in the parameters
of the fermion action
$m_0$ and $m_1$.  Integrating also over the (compact) link
variables of the gauge field leaves an unnormalized
expectation value which again is an analytic function of these
variables, including for $m_1=0$.   Similarly the
partition function $Z$ is strictly positive for a two-flavor
theory with Wilson fermions (for $g_0>0$), and again $Z^{-1}$ is bounded
for a bounded range of values of $m_0$ and $m_1$ which includes
$m_1=0$.  Hence normalized expectation values are bounded as well.
In contrast,
in quenched QCD the fermion determinant is missing, and a fermionic
observable may diverge if the Wilson--Dirac operator has near-zero
modes.  This is what happens in the case of the pion two-point
function.

It follows, through the Banks--Casher relation, that $\r(0)=0$.
Since $0 \le \r_l(0) \le \r(0)$, one has $\r_l(0)=0$ for any finite volume.
While of course the infinite volume is subtle (and does not commute with the
limit $m_1\to 0$), we believe that for any finite $l$,
$\r_l(0)$ will remain zero in the infinite-volume limit,
because $\r_l(0)$ reflects short-distance physics.
Also, if the inequality~(\ref{div}) remains valid in the infinite-volume
limit on the one hand,
and unquenched QCD does not have a $1/m_1$ divergence on the other hand,
clearly $\r_l(0)$ must remain zero in that limit.  The fact
that a $1/m_1$ divergence indeed cannot occur in unquenched QCD is
explained in the next subsection.

\vspace{5ex}
\noindent {\large\it 5.2~The Goldstone theorem}
\vspace{3ex}

Here we re-derive the Goldstone theorem in the euclidean path-integral
context in infinite volume (for a review see Ref.~\cite{GHK}).
For simplicity, we do this in the continuum; nothing relevant
changes on the lattice.  As in the previous subsection
our aim is to show that the (Fourier-transformed) pion two-point
function $\tG(p)$
is bounded in a neighborhood of $m_1 = 0$ but now for $p\ne 0$
(and without assuming that the condensate necessarily vanishes).
We may then again conclude that, for $p\ne 0$,
$2m_1 \tG(p) \to 0$ for $m_1\to 0$.  Taking this limit,
the Ward identity~(\ref{ppft})
yields (dropping terms of order $ap$)
\begin{equation}
  ip_\m \tG_\m(p)  = \svev{\p_3} \,, \qquad p \ne 0 \,.
\label{GB}
\end{equation}
If the condensate is non-zero (\ie\ if SSB takes place),
this implies the existence of a massless Goldstone pole in $\tG_\m(p)$.

To prove the boundedness of $\tG(p)$ for $p \ne 0$,
we invoke the K\"all\'en-Lehmann representation.
A one-particle contribution to $\tG(p)$ must be of the
form $|F|^2/(p^2+M^2)$ where $M$ is the particle's
mass and $F$ is a form factor which is non-zero on shell
($p^2=-M^2$). For any non-zero euclidean momentum,
\begin{equation}
  {m_1\over p^2+M^2} \le {m_1\over p^2}\,,
    \qquad p \ne 0\,.
\label{p2}
\end{equation}
The right-hand side of inequality~(\ref{p2})
provides a bound for the contribution of any one-particle
excitation to  $2m_1 \tG(p)$, implying that it vanishes
in the limit $m_1\to 0$, for $p\ne 0$.
This is true also if the particle's mass vanishes in the same limit
(as in the case of a Goldstone boson).
The contributions of multi-particle
states will be less infra-red singular, and should also vanish.
In summary, in unquenched QCD, $2m_1 \tG(p)$ vanishes in
the thermodynamical limit, for $p \ne 0$.

The quenched theory is not unitary, and the K\"all\'en-Lehmann
representation
for the pion two-point function invoked above is not valid.
Apparently, nothing stops the pion two-point function from developing a
$1/m_1$ divergence.

\vspace{5ex}
\noindent {\large\bf 6.~Implications for domain-wall and overlap
fermions}
\secteq{6}
\vspace{3ex}

The most common constructions of lattice Dirac operators with
domain-wall or overlap fermions employ the Wilson--Dirac operator
discussed in this paper (or improved versions thereof) as
a kernel.

Domain-wall fermions are five-dimensional Wilson fermions, in which
only hopping terms in the four physical directions couple
to four-dimensional link variables, which themselves
are independent of the
fifth coordinate \cite{dwf,fs}.  In the most common version, the fifth
dimension is restricted to a finite interval of length $a_5 N_s$,
where $a_5$ and $N_s$ are respectively the lattice spacing
and number of sites in the fifth direction.
Free boundary conditions are employed on either side (in the limit
of a vanishing physical quark mass).  In the limit
$N_s\to\infty$, massless four-dimensional fermions appear
which are bound to the two boundaries.  If the
left-handed component of this massless fermion is bound to one
boundary, the right-handed component is bound to the other
one.  For $g_0\to 0$, precisely one such massless fermion appears
(per five-dimensional fermion field) if the domain-wall height $M=-m_0$
is chosen such that $0<a'M<2$ where $a'=\max\{a,a_5\}$ \cite{dwq}.
This means that $m_0$ has to be super-critical.

The overlap-Dirac operator is defined as \cite{ovlp}
\begin{equation}
\label{Doverlap}
aD_{ov}= 1 - \g_5 \gh\,,\ \ \
\gh \equiv {H(m_0)\over |H(m_0)|}\,,
\end{equation}
with $H(m_0)$ the hermitian Wilson--Dirac operator.
Notice that $\gh^2=1$.
For this operator to describe one massless flavor in the
continuum limit, one needs to choose $0>am_0>-2$.

Because of the fact that $H(m_0)$ plays a crucial role
in the construction of domain-wall/overlap fermions, it is
natural to expect that the phase diagram of QCD with Wilson
fermions has important implications for properties of
domain-wall/overlap fermions. The most important dynamical issue
is whether the mobility edge is (close to) zero, or not.
In this section, we will
argue that in order to retain locality in lattice QCD with
overlap fermions, the parameters of the lattice theory must
be chosen well outside the Aoki phase.  For domain-wall
fermions the situation is equivalent: only well outside
the Aoki phase will chiral symmetry be restored exponentially
fast with increasing $N_s$.  In addition, only in that case will
the four-dimensional effective theory (which is described by a
generalized overlap operator, the details of which depend on $a_5$)
be local in the limit $N_s\to\infty$.

In the quenched case, we may say that lattice QCD with
domain-wall/overlap fermions
is inside or outside the Aoki phase if the ``underlying" theory
with the Wilson--Dirac operator $H(m_0)$ is.
One may think of the phase diagram
as that of a theory with $N_f$ quenched domain-wall/overlap
flavors {\it and} two quenched Wilson flavors.  Of course, this changes
nothing in the correlation functions of the domain-wall/overlap
quarks.  What it means is that,
also in the domain-wall/overlap case, we define the quenched Aoki phase
by the existence of a Goldstone pole in the appropriate correlation
function constructed from the inverse of the Wilson--Dirac operator $H(m_0)$
(\cf\ Eq.~(\ref{Gammas})), using  the same value of $m_0$
as in the kernel of the domain-wall/overlap operator,
and the same Boltzmann weight.
According to our conjecture, this corresponds to the region in
parameter space in which the mobility edge of $H(m_0)$ vanishes.

In the unquenched case, the only thing that changes is the
Boltzmann weight used to generate the
ensemble of gauge-field configurations on which the domain-wall
or overlap operator is computed.\footnote{
  Note that the fermion determinant used for generating the
  gauge-field configurations may correspond to any type of
  lattice fermion. If the sea quarks and the valence quarks
  are not coming from the same fermion action,
  one is dealing with a partially quenched theory \cite{pq}.
}
For any given ensemble, the mobility edge of the Wilson--Dirac kernel
$H(m_0)$ should have a well-defined value.
We may thus still define the Aoki phase as that region in
parameter space in which the mobility edge of $H(m_0)$ vanishes.
Conversely, if the mobility edge does not vanish, we will say
that we are outside the Aoki phase for the unquenched theory.

Let us start with locality of the overlap operator.
The overlap operator $D_{ov}(x,y)$ cannot have a finite range
\cite{horvath}, and thus is not strictly local
for a finite lattice spacing.  However,
Hern\'andez, Jansen and L\"uscher proved that
the overlap is local in the sense that
$|D_{ov}(x,y)|$ decays exponentially with the distance $|x-y|$,
provided the gauge field obeys an admissibility condition
\cite{hjl}.  The effect of this condition is to secure an $O(1/a)$ gap
in the spectrum of $H(m_0)$
(except very close to the critical values of $m_0$).
They furthermore generalized this result to the case that
$H(m_0)$ has an isolated zero mode inside an otherwise
empty spectral gap by showing that such a mode is necessarily
exponentially localized.  The rate of the exponential decay
of the overlap operator in these cases is of order one in
lattice units.

The problem is that, for realistic simulations,
it is impractical to impose an admissibility
condition on the gauge fields.  If instead one uses one
of the commonly used local gauge actions to generate an
equilibrium ensemble, $H(m_0)$ will have (localized)
eigenmodes with very small eigenvalues, and these eigenvalues will not be
isolated in the sense of Ref.~\cite{hjl}.
In fact, since the number of localized modes grows in proportion to
the volume, in the infinite-volume limit the
eigenvalues of the localized modes will form a dense set,
and no eigenvalue will be isolated.
This is most clear in the quenched case, where $\r(0)$ is always
non-zero in the super-critical region.

At this point we invoke our physical picture of the phase diagram.
For any $g_0>0$, the band edge $\l^0_{min}$ (of
the free hamiltonian $H_0(m_0)$) gets replaced
by the mobility edge $\l_c$. We hypothesize that, as long as $\l_c>0$,
the conclusion of Ref.~\cite{hjl} still holds:
if all near-zero modes are exponentially localized with
a finite average localization length $\bl$, then $\gh$ (and thus
$D_{ov}$) decays exponentially. As we argue below, the decay rate is
in principle governed by the smaller of $\l_c$ and $1/\bl$.  We expect this
situation to apply inside phase C in Fig.~1, well
outside the Aoki phase.  In fact, if we assume, like at the
end of Sect.~4.2, that the near-zero modes have a finite
maximal localization length $l_{max}$
outside the Aoki phase, our hypothesis is simple to prove.

First, however, let us consider the corresponding situation
with domain-wall fermions.
The relevant Ward identity~\cite{fs} for the study of
the finite-$N_s$ chiral symmetry breaking
with domain-wall fermions is, for  $x\ne y$,
\begin{equation}
   \db_\m \vev{A^a_\m(x)\, J^b_5(y) } =
   2m_q \vev{J^a_5(x)\, J^b_5(y) }
   + 2 \vev{J^a_{5q}(x)\, J^b_5(y) }\,.
\label{pionms}
\end{equation}
Here $A^a_\m$ is the domain-wall partially-conserved axial current and
$J^b_5$ is the corresponding pseudo-scalar quark density
($a,b$ label flavor generators); $m_q$ is the quark mass.
$J^a_{5q}$ is a pseudo-scalar density located midway between the
boundaries of the five-dimensional bulk.  The term containing
this density represents the chiral symmetry breaking at finite $N_s$ other
than the expected breaking coming from an explicit quark mass.  For chiral
symmetry to be restored, it should vanish in the $N_s\to\infty$ limit
for non-singlet axial currents.  (In the case of the axial U(1)
current this term gives rise to the anomaly in the continuum limit.)

In order to study chiral symmetry restoration, it is useful to
consider the transfer matrix $T(M,a_5)$ which hops in the
fifth direction, from one four-dimensional ``time" slice
to the next \cite{fs,oovlp}.
For every eigenmode $\J_n$ of $T(M,a_5)$ with
(positive\footnote{
  For a certain range of $M$ and $a_5$ the transfer matrix may have
  real but negative eigenvalues; in this case the entire analysis
  can be carried out in terms of the transfer-matrix squared
  (provided $N_s$ is a multiple of four).
})
eigenvalue $\o_n$, we let $q_n = \min\{\o_n,\o_n^{-1}\}$.
Hence, $0<q_n\le 1$. In the {\it second-quantized} transfer matrix,
the replacement of $\o_n$ by $q_n$ reflects normal ordering.
For the free theory the transfer matrix has a gap:
$q_n \le q_0$ for all eigenmodes, where $q_0<1$.
We refer to $q_0$ as the band edge of the free transfer matrix.
(As an example, in the special case where $aM=a_5/a=1$,
one has $q_0=1/2$.)
In the interacting theory, we define
the mobility edge of the transfer matrix as $q_c=\max\{q_n\}$,
where the maximum is taken over the extended modes only.
Below, we will also speak of the ``hamiltonian''
$H(-M,a_5)\equiv-\log{(T(M,a_5))}/a_5$.
This hamiltonian has a band edge $-\log(q_0)/a_5$ in the free theory,
and a mobility edge $\l'_c = -\log(q_c)/a_5$ in the interacting theory.
(In the limit $a_5\to 0$ one recovers the familiar Wilson--Dirac operator:
$H(-M,0)=H(-M)$, and $\l'_c$ reduces to $\l_c$.)

Since the transfer matrix has a gap in the free theory,
the four-dimensional massless fermions at the boundaries
are exponentially bound to their respective boundaries.
If we take $N_s\to\infty$ at fixed $M$ and $a_5$,
we find an exponentially decreasing overlap
between the (fifth-coordinate) wave functions of the light-quark modes
tied to the boundaries and the ``midway'' pseudo-scalar density $J^a_{5q}$.
Hence chiral symmetry gets restored exponentially in $N_s$.
With gauge fields obeying an admissibility condition,
the situation remains the same, because the transfer matrix still has a gap.
Also, with commonly used local gauge actions,
a similar behavior has been demonstrated in weak-coupling
perturbation theory \cite{dwpt,yspt}.

Non-perturbatively, one can prove that, for any $a_5$,
a zero mode of the Wilson kernel is also an eigenstate with eigenvalue one
of the transfer matrix: $T(M,a_5)\J_n=\J_n$ if and only if $H(-M)\J_n=0$
\cite{fs,oovlp}. This implies that $H(-M)$ and $H(-M,a_5)$
have a similar spectrum of near-zero modes.
A near-zero mode of $H(-M,a_5)$ causes long-range
correlations in the $s$ direction.
This means that in realistic simulations
with commonly used gauge actions,
configurations will occur for which the last term in Eq.~(\ref{pionms})
may be large, thus ``threatening" the chiral
symmetry of domain-wall fermions even in the large-$N_s$ limit.
This danger is again brought out
most clearly in the quenched case, where a non-zero density of
near-zero modes always occurs for the above specified range of $M$.

However, we hypothesize that, in analogy with the overlap case,
there is a fundamental difference between the effect of
exponentially localized near-zero modes and that of extended ones.
If all near-zero modes of $H(-M,a_5)$ are
exponentially localized with an average localization length $\bl$,
we expect their contribution to the symmetry-breaking term
in Eq.~(\ref{pionms}) to decay exponentially as a function of the
{\it four-dimensional} separation $|x-y|$.
If the scale of $\bl$ is set by the (four-dimensional)
lattice spacing, this decay will resemble
a contribution from excitations with mass of the order of the cutoff.
When the lattice spacing is small enough,
even at finite $N_s$ this contribution will vanish for large $|x-y|$
relative to the other two terms in Eq.~(\ref{pionms}), whose
long-distance behavior is determined by a {\it physical} mass.
All other contributions to the symmetry-breaking
term will not be suppressed with the four-dimensional separation,
but they will vanish exponentially with $N_s$.

Similar conclusions apply to the locality of the
effective four-dimensional lattice Dirac operator in the $N_s\to\infty$
limit~\cite{effD}. This is a generalized overlap operator constructed
by making the replacement $H(-M) \to H(-M,a_5)$ in Eq.~(\ref{Doverlap}).
Because of the similar zero-mode structure of $H(-M)$ and $H(-M,a_5)$,
we expect the effective Dirac operator to be local
(in the exponential sense) well outside
the Aoki phase, as in the case of the overlap discussed above.

For both domain-wall and overlap fermions,
these arguments break down if the near-zero modes are
extended, or even, if $1/\bl$ becomes of the same order as the
numerical values of the physical masses in a particular
simulation.  This latter situation is expected close to the Aoki
phase, because $\bl$ will increase going towards the phase
transition, and become infinite at the phase transition.
The domain-wall formalism gives a clear indication on what may go
wrong inside the Aoki phase. When the mobility edge of the Wilson kernel
$H(-M)$ (or, more generally, of $H(-M,a_5)$)
is zero, there are massless excitations everywhere
{\it inside} the five-dimensional bulk.
Normally,  the contribution of the {\it massive} bulk
modes is canceled by Pauli-Villars (pseudo-fermion) fields.
But there is absolutely
no guarantee that that cancellation will persist when
the bulk fermion and pseudo-fermion modes are massless:
this situation corresponds to a different phase of the theory.
The limit $N_s\to\infty$ now involves infinitely many
unphysical, light four-dimensional fields
(arising from both the five-dimensional
domain-wall and the pseudo-fermion fields),
and we have every reason to worry that the limiting theory
is not what we want it to be.  This is true for any $a_5$, and thus also
in the overlap limit $a_5\to 0$.

We now present a more detailed argument as to why
the overlap operator should be local well outside the Aoki phase.
We will assume, as we did at the end of Sect.~4.2, that at
a given point well outside the Aoki phase there exists a finite
maximum localization length for the near-zero modes of $H(m_0)$.
In fact, to avoid technical complications, it is convenient to make
a slightly stronger assumption, namely, that if $\l_c>0$ is the mobility edge,
then all eigenstates with $-\l_c/2<\l<\l_c/2$ have a bounded localization
length $l\le l_{max}<\infty$.
As before, this assumption is not unreasonable
because the average localization length diverges only for
$|\l|\nearrow \l_c$, and the probability for arbitrarily large localization
lengths is expected to vanish rapidly
(and uniformly) for any $|\l|<\l_c/2$.
By simply replacing $H(m_0)$ by $H(-M,a_5)$,
the below argument may also be applied to the $N_s\to\infty$ limit of
domain-wall fermions.\footnote{
  For small $a_5$ one can also use Borici's kernel \cite{bor}
  $\g_5 D(-M)(2-a_5 D(-M))^{-1}$
  that gives rise to the same
  generalized overlap operator as $H(-M,a_5)$.
}

We begin by noting that
the overlap operator $D_{ov}$ is local if $\gh$ is local.
We split
\begin{equation}
  \gh=\gh^< + \gh^> \,,
\label{g5split}
\end{equation}
where $\gh^<$ is the projection of $\gh$ onto the subspace
spanned by eigenstates with eigenvalue $|\l|<\l_c/2$.
On the basis of eigenmodes of $H(m_0)$, $\gh^<$ can be represented
as
\begin{equation}
  \gh^<(x,y)
  = a^4 \sum_{|\l_n|<\l_c/2} \J_n(x) {\l_n\over |\l_n|} \J^\dagger_n(y) \,.
\label{specg}
\end{equation}
Assuming that all modes with $|\l|<\l_c/2$
satisfy inequality~(\ref{lclz}) one has
\begin{equation}
  |\gh^<(x,y)|
  \le
  c_1 a^4 \sum_{|\l_n|<\l_c/2}\;
  {1 \over l_n^4}\; \exp\left(-{|x-x_n^0|+|y-x_n^0|\over 2 l_n}\right) \,.
\label{xyRR}
\end{equation}
Performing the ensemble average as in previous sections (see in
particular Eqs.~(\ref{limRdef}) and~(\ref{intll}))
we find
\begin{equation}
  \Big\langle|\gh^<(x,y)|\Big\rangle
  \le
  c_1 a^4 \int_a^\infty dl\,
  \int_{-\l_c/2}^{\l_c/2} d\l\,\r'_l(\l) K(|x-y|/l) \,,
\label{ghvev}
\end{equation}
where
\begin{equation}
  K(|x-y|/l)
  = l^{-4} \int d^4x_0\,
  e^{-(|x-x_0|+|y-x_0|)/(2 l)}
  \approx {|x-y|\over l} \, e^{-|x-y|/(2 l)}\,.
\label{K}
\end{equation}
The last approximate equality holds for $|x-y| \gg l$,
where the integral is dominated by localized modes
supported inside a tube of radius $l$
around the straight line connecting $x$ and $y$.
We now invoke the assumption that $l_n \le l_{max}$
for all  $-\l_c/2<\l_n<\l_c/2$. The ``worst-case scenario'' is that
the $l$-integral in Eq.~(\ref{ghvev}) will be dominated by $l \leqx l_{max}$.
In this case we find the bound
\begin{equation}
  \Big\langle|\gh^<(x,y)|\Big\rangle
  \leqx
  \exp(-|x-y|/(2 l_{max})) \,,
\label{projRR}
\end{equation}
where constants and power corrections have been ignored.
Hence $\gh^<$ is local.

The operator $\gh^>$ lives in the orthogonal subspace,
whose projection operator is
$\cp_>(x,y) \equiv \d_{x,y} -\cp_<(x,y)$.
Here
\begin{equation}
\label{plcl}
\cp_<(x,y) = a^4 \sum_{|\l_n|<\l_c/2}\J_n(x)\J^\dagger_n(y)
=(\gh^<)^2\,,
\end{equation}
is the projector on the eigenstates with $|\l|<\l_c/2$.
The bound~(\ref{xyRR}) evidently applies to $\cp_<(x,y)$ as well.
Therefore $\cp_<$ is local, and, hence, also $\cp_>$ is local.
Proceeding exactly as in Ref.~\cite{hjl} we write
\begin{equation}
  \gh^> = H(m_0)\, |H(m_0)|^{-1}\, \cp_> \,.
\label{rest}
\end{equation}
(Strictly speaking, the right-hand side is defined via its mode expansion.)
The expansion of (the eigenvalues of) $|H(m_0)|^{-1}$
in terms of Legendre polynomials
may now be invoked, and the locality of $\gh^>$ follows.

Let us discuss some implications of this result.
In principle one can envisage two extreme situations.
If $l_{max} \ll \l_c^{-1}$ then the non-locality of $\gh^<$ and $\cp_>$
may be neglected. The localization range of the overlap will be governed by
the mobility edge $\l_c$, and will be of order $\l_c^{-1}$.
In other words, in the absence of an admissibility constraint,
$\l_c^2$ plays the role of the lower bound on the spectrum of $H^2(m_0)$.
The opposite extreme is that $l_{max} \gg \l_c^{-1}$, where we expect
the localization range of the overlap to coincide with $l_{max}$.

In our argument, we did not attempt to maintain the same
level of rigor as in Ref.~\cite{hjl}.
The advantage of our more heuristic argument is that it deals with
the more realistic case of a {\it density} of localized
near-zero modes where
the methods of Ref.~\cite{hjl} are inapplicable.
Our analysis is only semi-realistic because, in principle,
there will be a
very tiny, but non-zero probability to encounter an
exponentially localized near-zero eigenmode with an
arbitrarily large localization length (in a large enough
volume).  In practice, however, we expect that our assumption of the
existence of a  finite, not too large
$l_{max}$ will be valid for most simulations.
Of course, the most important question in any given
simulation is how the scale of the typical localization length
as well as the size of the mobility edge
compare quantitatively to the scale of the physics one is
trying to compute.

We now turn to domain-wall fermions with finite $N_s$.
A useful measure of chiral symmetry breaking
is the so-called residual mass $\mres$ \cite{rm},
which is essentially the ratio of the two correlators on the right-hand side
of Eq.~(\ref{pionms}). Denoting the (euclidean) time coordinate by $\t$ we define
\begin{equation}
  \mres(\t,N_s) =
       {\sum_{\vec{x}}\vev{J^+_{5q}(\vec{x},\t)\, J^-_5(\vec{0},0) }
            \over
      \sum_{\vec{x}}\vev{J^+_5(\vec{x},\t)\, J^-_5(\vec{0},0) }}\,,
\label{mres}
\end{equation}
where for convenience we have switched to flavor-changing densities.
(In terms of the ``isospin'' symmetry of the two flavors occurring
in these densities,
the $\pm$ superscripts correspond to the operators $\t_\pm$ of Eq.~(\ref{ISO}).)

The study of $\mres$ consists of two steps.
First,  using the transfer matrix formalism of Ref.~\cite{fs}
we derive expressions for the correlators in a fixed gauge-field background.
For the numerator in Eq.~(\ref{mres}) this step is outlined in Appendix~B.
Next we have to carry out the integration over the gauge field.
In general this is quite complicated, and so
we will restrict ourselves to two relatively simple cases. Observing that
$\t$ is a measure of the four-dimensional separation in Eq.~(\ref{mres}),
we will discuss
the $\t$ dependence for fixed $N_s\gg 1$, and
the $N_s$ dependence for fixed $|\t|\gg a$.
As before, well inside the C phase we expect the mobility edge of $H(-M,a_5)$
to be a quantity of order one in lattice units.
For $\t=O(a)$ and $N_s\gg 1$, $\mres$ will be dominated by
the exponentially localized near-zero modes,
whereas for $N_s=O(1)$ and $|\t| \gg a$ it will be dominated by
the extended states close to the mobility edge.

The crucial observation is that, as explained earlier,
the quark and anti-quark operators of the pseudo-scalar density
$J^-_5(\vec{0},0)$ live on the two boundaries of the fifth dimension,
whereas $J^+_{5q}(\vec{x},\t)$ is located midway between the two boundaries.
In a diagrammatic language,
the fermion (anti-fermion) at the boundary must be connected by a fermion line
to the ``midway'' anti-fermion (fermion). In  Eq.~(\ref{TOmRR}),
these two fermion lines come from the one-particle sector
of the second-quantized transfer matrix that acts $N_s/2$ times
on the fermion (or anti-fermion) at the boundary.
The contribution of each mode to a fermion line involves a factor of
$\J_n(\vec{0},0)\, q_n^{N_s/2}\,\J_n^\dagger(\vec{x},\t)$,
or its hermitian conjugate.

For $|\t| \geqx a$ the denominator in Eq.~(\ref{mres}) will contain
short-distance contributions which are of no interest to us.
In this case it is simpler to consider the numerator alone.
Assuming some fixed $N_s\gg 1$, the contribution of all extended modes
has died out, because for them, $q_n^{N_s/2} \leqx q_c^{N_s/2} \ll 1$.
We are left with the contribution of the exponentially localized modes
of $H(-M,a_5)$, of which the near-zero modes dominate for $N_s\gg 1$.
Following closely the overlap case,
Eq.~(\ref{lclz}) allows us to put a bound on
$\J_n(\vec{0},0)\,\J_n^\dagger(\vec{x},\t)$.
A similar bound applies to the contribution of the second fermion line.
The product of the two bounds gives rise to the same exponential factor
as on the right-hand side of Eq.~(\ref{xyRR}),
except that the factor of $1/2$ inside the exponent is missing.
As before we now assume the existence of
a maximal localization length for the near-zero modes, arriving at
\begin{equation}
  \sum_{\vec{x}}\vev{J^+_{5q}(\vec{x},\t)\, J^-_5(\vec{0},0) }
  \sim
  \exp(-|\t|/l_{max}) \,, \qquad N_s \gg 1 \,.
\label{mrestau}
\end{equation}
The evident analogy with Eq.~(\ref{projRR}) is not surprising.

Finally, we consider the $N_s$ dependence for $|\t| \gg a$.
Returning to $\mres$,
now all the exponentially-localized modes may be neglected. We expect that
$\mres$ will be dominated by the extended modes of the transfer matrix
with eigenvalues close to the mobility edge.
The anticipated result is
\begin{equation}
  \mres \sim q_c^{N_s} \,, \qquad |\t| \gg a \,.
\label{mresNs}
\end{equation}
This reflects a cancellation of the $\t$ dependence
between the numerator and the denominator in Eq.~(\ref{mres}),
for large $\t$ separations
(for numerical evidence supporting this, see Ref.~\cite{rm}).
We are unable to derive this result analytically.
We will give a heuristic argument, based on an analogy with
perturbation theory, which supports this result.

In perturbation theory, this result is obtained as follows~\cite{yspt}.
When the quark mass is zero, the one-loop domain-wall fermion
propagator has, near each boundary, a factorizable form.
Assuming for definiteness that the right-handed quark resides
near the $s=0$ boundary, the propagator in the vicinity of this boundary is
\begin{equation}
  G(x,y;s,s') \approx \, \c(s)\, P_R\, G(x,y)\, \c^\dagger(s') \,.
\label{fctrz}
\end{equation}
Here $P_R=\half(1+\g_5)$,
$x,y$ are the usual four-dimensional coordinates, and
$s,s'$ are coordinates in the fifth dimension.
The separation $|x-y|$ is assumed to be large compared to the lattice scale
(but small enough for weak-coupling perturbation theory to be applicable).
$G(x,y)$ is the effective quark propagator whose tree-level Fourier transform
is $1/\sl{p}$ for small $p$, and which, at higher orders,
contains the familiar logarithmic self-energy corrections of a massless quark.
The $s$-coordinate wave function $\c(s)$ carries no Dirac indices.

For an optimally-chosen $M$ as a function of $g_0$,
the (tadpole-improved) tree-level wave function $\c_0(s)$ is completely
confined to the boundary layer $s=0$. For $s \ge 1$,
the wave function arises from a short-distance, one-loop {\it quantum} effect.
Explicitly, $\c(s) = \c_0(s) + g_0^2\; \c_{quantum}(s)$
where $\c_0(s)=\d_{s,0}$ and
$\c_{quantum}(s) \sim q_0^s$ up to a power correction.
This form of $\c_{quantum}(s)$ gives rise to Eq.~(\ref{mresNs}), with $q_c \to q_0$.
(Since $q_0$ is the band edge of the free transfer matrix,\footnote{
  In Ref.~\cite{yspt}, the band edge of the free transfer matrix
  was denoted $q_1$ (and not $q_0$).
}
the shape of the leading-order $\c_{quantum}(s)$ depends
on the free domain-wall fermion action only.
In higher orders one expects
$\c_{quantum}(s) \sim (q_0 f(g_0))^s$
where $f(g_0)=1+O(g_0^2)$, and where $f(g_0)$ depends on the gauge action too.)

Restoring the momentum dependence, the $s$-coordinate wave function
$\c(s;p)$ is universal in the sense that
corrections to $\c(s)=\c(s;0)$ vanish like $(ap)^2$. The physical reason
is that the propagation in the fifth direction is dominated
by a small region in the Brillouin zone surrounding the critical
momenta that saturate the band edge $q_0$. These momenta are $O(1/a)$.
Fluctuations of the gauge field allow low-momentum
modes to couple to these lattice-scale modes,
with an amplitude that can be naturally expanded in powers of $(ap)^2$.
Eq.~(\ref{fctrz}) is obtained by keeping only the leading term in this expansion.
For all other terms, the $1/\sl{p}$ singularity of the (free)
propagator is wiped out, and so their effect is negligible
for large space-time separations.

Non-perturbatively, we may envisage a similar factorization
of the domain-wall fermion propagator for $|\t| \gg a$
(in a fixed gauge).
Again, the extended states close to the mobility edge of the transfer
matrix will mediate the $s$-propagation.
Again these extended states will be controlled by the lattice scale,
and it is reasonable that their coupling to low-momentum external states
will have a universal value.
Turning to the correlators of the gauge-invariant pseudo-scalar densities,
a factorizable form of the (fixed-gauge) domain-wall fermion propagator,
as in Eq.~(\ref{fctrz}), implies that the common $\t$ dependence of
the numerator and the denominator of Eq.~(\ref{mres}) is $\exp(-|\t|m_\p)$.
The result is Eq.~(\ref{mresNs}).\footnote{
  For the DBW2 gauge action at quenched $a^{-1}\sim 2$~GeV
  it was found that $q_c \sim 0.6$ \cite{Db}. This is very
  close to the value $q_0=0.5$ found in one-loop
  perturbation theory \cite{yspt}. Therefore, in this case both
  higher-order effects and non-perturbative effects are small.
}

\vspace{5ex}
\noindent {\large\bf 7.~Conclusions}
\secteq{7}
\vspace{3ex}

Let us summarize our conclusions and make some
additional comments.  We start with what we learned about the
quenched phase diagram.

In finite volume, we argued that the Aoki condensate is non-zero
everywhere in the super-critical region of the phase diagram,
and that the pion two-point function always has a $1/m_1$
divergence.  This  $1/m_1$ divergence arises mainly from
exponentially localized near-zero modes.
If the restricted spectral density $\r_l(0)$ is non-zero
for some localization length $l$,
then the momentum-space pion two-point function $\tG(p)$
exhibits the $1/m_1$ divergence for all momenta up to $p^2 l^2 \sim 1$.
Because of this divergence, clearly the finite-volume quenched
theory is only well defined with a non-vanishing twisted mass
$m_1$.

Extending this to infinite volume, we argued moreover that if all
near-zero modes are localized, the condensate is approximately equal
to $\lim_{m_1\to 0}2 m_1 \tG(p)$, and the difference vanishes with $p^2$.
This implies that there are no Goldstone excitations.

Adopting the mobility-edge hypothesis
we arrive at the following physical picture.
In the super-critical quenched theory the pionic
condensate is always non-zero.
Goldstone poles exist, however, only in part of the super-critical region
which, by definition, is the quenched Aoki phase.
Inside the Aoki phase the mobility edge vanishes,
all the near-zero modes are extended,
and the pion two-point function has no $1/m_1$ divergence,
because the contribution of extended modes to this divergence
goes to zero in the infinite-volume limit.
Outside of the Aoki phase the mobility edge is larger than zero,
and all the near-zero modes are exponentially localized.
There are no Goldstone poles, and the Ward
identity~(\ref{pp}) is saturated instead by
the $1/m_1$ divergence of the pion two-point function.

We can, in fact, completely characterize the quenched
phase diagram using {\it two} local order parameters.
For the SU(2)-orientation of the twisted-mass term in the
action~(\ref{L}),
one order parameter is the usual condensate $\svev{\p_3}=2\p\r(0)$.
The other is $\x$, defined in Eq.~(\ref{xi}), which measures the
size of the $1/m_1$ divergence in the pion two-point function.
In the quenched theory $\svev{\p_3} \ne 0$, and SU(2) is spontaneously
broken, in the entire super-critical region.
Hence the lines $m_0=0$ and $am_0=-8$ are phase boundaries
in the quenched theory.  Inside the super-critical region we have
a phase with Goldstone bosons (the Aoki phase),
and phases with no Goldstone bosons
(the C phases, and the super-critical parts of the A regions,
which form separate phases in the quenched theory).
The order parameter for having {\it no} Goldstone excitations
is $\x>0$.

In continuum, infinite-volume, quenched chiral perturbation theory,
it is known that the usual condensate $\svev{\bj\j}$ develops a
$\log(m)$,
or $m^{-\d}$ with $\d$ small, divergence in the limit $m\to 0$
\cite{bg,sh}.  In the continuum limit of the lattice theory
chiral symmetry is restored, and the Aoki condensate can be rotated back
to the usual condensate (see for example Ref.~\cite{shsi}).
Turning this argument around, one expects a $\log(m_1)$
divergence of the Aoki condensate inside the Aoki phase
in the continuum limit,
with $O(am_1)$ corrections due to a non-zero lattice spacing.
As explained above, we expect no $1/m_1$ divergence of the pion
two-point function inside the Aoki phase in the infinite-volume limit.

In unquenched QCD we do not have any new results.
We did verify that, {\it only} in unquenched QCD, one can
prove that the pion two-point function $\tG(p)$ is bounded.
In finite volume this is always true, while
in infinite volume this is true provided $p \ne 0$.
Thus, our new results for the quenched theory
do not contradict any of the well-known facts for the unquenched
case.

The main differences between unquenched and quenched QCD are as follows.
Inside the Aoki phase, the differences are accounted for by the chiral
lagrangian, and the quenched theory is known to have enhanced chiral
logarithms.
Outside the Aoki phase, both unquenched and quenched QCD have
exponentially localized near-zero modes;
in the infinite-volume limit, zero is not an isolated eigenvalue
of any gauge-field configuration drawn according to the Boltzmann weight.
However, in the quenched theory $\r(0)\ne 0$,
whereas in the full theory $\r(0)=0$.
Heuristically, this can be understood as follows.
Let us compare two infinite-volume gauge-field configurations,
which differ only inside a small region
whose radius is $O(1)$ in lattice units.
Assume that, inside this small region,
each configuration supports an exponentially localized mode, with
possibly different eigenvalues.
In the quenched theory, these two configurations can have
similar Boltzmann weights; but in the unquenched two-flavor theory,
each localized mode will contribute to
the Boltzmann weight of the corresponding configuration
(and, hence, to the spectral density)
a factor of $\l^2+m_1^2$ (\cf\ Eq.~(\ref{detF})).
Therefore the configuration with the smaller
eigenvalue will be suppressed.
In the limit $m_1\to 0$, a configuration supporting an
exponentially localized
zero mode is completely suppressed in full QCD.
In this argument, it is important that the difference between
the two configurations is confined to a small region.
If all the near-zero modes
are extended, a local change in the configuration
will have only an infinitesimal effect on any one of the eigenvalues,
and the entropy may be large enough to overcome the suppression
factor due to the fermion determinant.  Thus, the extended modes
may still build up a condensate, despite the suppression by the
fermion determinant.

On the basis of our analysis, we argued that the domain-wall
and overlap formulations work only inside the C phase(s).\footnote{
  One quark field per one lattice-fermion field is obtained
  by taking the continuum limit inside the C phase that borders on
  the interval $-2<am_0<0$ on the $g_0=0$ axis
  (where $M=-m_0$ is the domain-wall height,
  and provided $a_5 \le 1$ \cite{dwq}). A possible trajectory
  for taking the continuum limit
  is indicated by the dashed line in Fig.~1.
}
Inside those phases the mobility edge is non-zero,
and all the near-zero modes of the Wilson kernel
are exponentially localized.  The overlap should be local, and
for domain-wall fermions, chiral symmetry should be recovered
exponentially with the size of the fifth dimension.
Also, the effective four-dimensional overlap operator
emerging in the $N_s\to\infty$ limit (which in the
$a_5\to 0$ limit coincides with the standard overlap constructed
with a Wilson-fermion kernel) should be local.
Inside the Aoki phase one encounters infinitely many light unphysical
modes, which contribute to the logarithm of the partition function
with opposite signs.  The overlap operator corresponds to the special
case $a_5\to 0$.  It is hard to imagine how either formulation
could remain valid under these circumstances.\footnote{
  For other work pointing at difficulties at strong coupling,
  see Ref.~\cite{benetal}. For related work on the phase structure of
  overlap fermions with a small hopping parameter, see Ref.~\cite{hopp}.
}

It is worth commenting on the role of the gauge action in this
respect.  In particular, let us discuss how precisely an
admissibility condition \cite{hjl} would change the picture.
An admissibility condition means that all the plaquette
variables are constrained to be closer to one than some $0<\e \ll 1$.
The spectrum of $H^2(m_0)$ then has a lower bound
$(\l^0_{min})^2 - \d^2$ (again $\l^0_{min}$ refers to the free
hamiltonian $H_0(m_0)$), where $\d$ is determined in terms of $\e$.
This prevents the existence of
near-zero modes when $\l^0_{min} >\d$, but not when $\l^0_{min} < \d$.
The super-critical region will now be defined as that region in
which $H(m_0)$ has zero modes when restricted to gauge fields
satisfying the admissibility condition.  This new super-critical
region consists of five (disconnected) vertical strips
in the phase diagram, each of which is located near one of the five
critical value of $m_0$.
Inside each of those super-critical strips
there should still be an Aoki phase.

It is clear that, in principle, an admissibility condition
guarantees the locality of the (ordinary or generalized) overlap,
and the exponential recovery with $N_s$ of the domain-wall's chiral symmetry,
for $m_0$ not close to a critical value.
However, imposing an admissibility condition in numerical simulations
is prohibitively expensive.
Very similar results can however be achieved by modifying the
lattice gauge and/or fermion action,
such that the density of the localized near-zero modes
of $H(m_0)$ decreases for a fixed value of the lattice spacing.
The density of near-zero modes was studied directly
and indirectly in numerous publications;
for recent reviews see Ref.~\cite{PH}.
In an approximation where all the exponentially localized modes
are neglected, the mobility edge turns into a gap,
which can be studied in perturbation theory~\cite{yspt}
(see also Ref.~\cite{aokidw}).  Non-perturbatively,
numerical results with the Iwasaki~\cite{Iw} and DBW2~\cite{Db} gauge
actions show a dramatic depletion of near-zero modes.
This likely corresponds to larger
values for the mobility edge at fixed lattice spacing.
If we would replace the vertical axis in Fig.~1 by the lattice spacing
(in physical units) this would amount to a recess of the Aoki-phase
boundaries towards larger $a$, \ie\ to enlarged C phases.

We have studied the PCAC (Partially-Conserved Axial Current)
relation for domain-wall fermions, Eq.~(\ref{pionms}).
The residual mass defined in Eq.~(\ref{mres}) directly provides information
on the rate of chiral symmetry restoration as a function of $N_s$.
We argued that,
by monitoring the anomalous term in the domain-wall PCAC relation,
one can determine {\it two} crucial features of the ``hamiltonian''
$H(-M,a_5)$ that controls the propagation in the $s$-direction:
the mobility edge $\l'_c$ (equivalently $q_c$), \cf\ Eq.~(\ref{mresNs}),
and (in effect) the average localization length $\bl$ of the near-zero modes,
\cf\ Eq.~(\ref{mrestau}).
These results may be combined as
\begin{equation}
  \mres \sim {e^{-|\t|/\bl}\over f(\t)} + c\, q_c^{N_s} \,,
\label{mrestNs}
\end{equation}
where $c$ is a constant.
(Here we have traded the maximal localization length $l_{max}$
of the near-zero modes, assumed in Sect.~6,
by an average localization length $\bl$.)
The function $f(\t)$ contains the $\t$ dependence of the
domain-wall's pion two-point function
in the denominator in Eq.~(\ref{mres}).\footnote{
  As explained in sect.~6, for $|\t| \geqx a$
  Eq.~(\ref{mrestNs}) is not directly amenable to numerical tests
  because of the unknown short-distance effects in the pion two-point
  function; it may be advantageous to study to numerator of Eq.~(\ref{mres})
  directly, \cf\ Eq.~(\ref{mrestau}).
}
This result should be valid provided $N_s \gg 1$ and/or $|\t|\gg a$.

The residual mass has been extensively studied
in domain-wall fermion simulations which according to our terminology
correspond to the range $|\t|\gg a$.\footnote{
  The simulations often show a nice plateau
  $\mres(\t,N_s) \approx \mres(N_s)$
  for a wide range of $\t$.
}
As a function of $N_s$,
the results for the Wilson gauge action at quenched $\b=6.0$
and optimally chosen $M$ \cite{rm}
are characterized by a rapid initial drop of $\mres$.
However, for larger values of $N_s$
the fall-off slows down, and eventually $\mres$ settles at a non-zero value.
Therefore, the first term on the right-hand side
of Eq.~(\ref{mrestNs}) is non-negligible in these simulations.
According to our physical picture, this suggests that
both the density of near-zero modes and their average localization
length may be relatively large; hence, in this case, one is very
close to the Aoki phase.\footnote{
  We believe that the first exploratory finite-temperature domain-wall
  simulations~\cite{ftDW} were in fact carried out inside the
  Aoki phase, which would explain the poor chiral symmetry observed even
  at relatively large $N_s$.
  This is why, in Fig.~1, we have drawn the region where domain-wall
  simulations have been tried such that it partially overlaps with
  the Aoki phase.
}
In contrast, the Iwasaki \cite{Iw} and DBW2 \cite{Db} results
exhibit a clean exponential fall-off for all values of $N_s$
where data are available. This implies that the effect of
exponentially localized near-zero modes vanishes within numerical accuracy.
In these cases the density of near-zero modes
should be very small, and their average localization length
should be of order one in lattice units; hence one is
well outside the Aoki phase.

The ``hamiltonian'' $H(-M,a_5)$ also serves as a kernel
of the generalized overlap operator obtained in the $N_s\to\infty$ limit,
and $\l'_c$ and $\bl$ are the two quantities which control the
locality of this operator.
The small but non-zero value of $\mres$, found using the Wilson gauge action
at quenched $\b=6.0$ when both $N_s$ and the separation are large,
suggests that in this case
the localization scale of the limiting operator might not be
sufficiently small compared to the physical scale of the simulation.
In contrast, the Iwasaki and DBW2 results imply a highly
localized effective four-dimensional operator.

When overlap fermions are employed, as with domain-wall fermions
it is necessary to verify whether locality of the overlap operator
is obtained, and whether the localization scale is small compared
to the physical length scale.
This should become a routine practice in any overlap simulation!

The picture for the phase diagram of
QCD with Wilson fermions we painted in this paper is
for a good part conjectural.  In particular, we did not prove
the hypothesis about the mobility edge presented in Sect.~4.2.
In addition, some of our more rigorous arguments are based
on the assumption that, in the regions where the mobility edge
does not vanish, a maximum localization length $l_{max}$
exists.  However, we would like to emphasize again that
our conjecture appears to be the simplest possible way in
which we can understand the collected evidence about the
quenched and unquenched phase diagrams, incorporating both
numerical \cite{Aokiq,Aokif,scri,jansenetal}
and analytical \cite{aoki,BNN,creutz,shsi,GSS} results.

In order to test our proposal for the phase diagram, it would
be interesting to study the Ward identity~(\ref{pp}) numerically,
in particular in the quenched two-flavor theory.  The second
term on the left-hand side should show the $1/m_1$ divergence
already in finite volume.  A comparison with the pionic
condensate will test whether the $1/m_1$
divergence saturates the Ward identity for very small
momenta.  A study of the dependence of all terms in the Ward
identity on the volume should make it possible to see in which
region of the phase diagram Goldstone excitations occur in the
infinite-volume limit.  Obviously, such a numerical study
will have to include a twisted mass \cite{tm}, and no (exceptional)
configurations should be discarded.  Numerical studies
can be extended to study localization lengths of near-zero
modes as well as non-zero modes outside the Aoki phase,
once the location of the various regions in Fig.~1 has been
established.  The localization length of non-zero modes would yield
information on the value of the mobility edge.
(As explained earlier, similar information may be obtained
from (quenched) domain-wall fermion simulations.) Similar studies
can also be done in the unquenched theory, but in that case
the spectral density of the localized modes will depend in a more
complicated manner on the randomness of the gauge field
because of the back-reaction through the fermion determinant.
It is interesting that the quenched theory provides a conceptually
simpler arena to test the validity of our conjecture!

\vspace{5ex}
\noindent {\large\bf Acknowledgements}
\vspace{3ex}

The topics of this work were addressed in the
RIKEN BNL Research Center workshop on {\it Fermion Frontiers in
Vector Lattice Gauge Theories}, held on May 6-9, 1998,
at Brookhaven National Lab.,
and more recently in the program
{\it Lattice QCD and Hadron Phenomenology},
held during September - December 2001
at the Institute for Nuclear Theory at the University of Washington, Seattle.
We thank the participants for extensive discussions which motivated this work.
YS also thanks Amnon Aharony, Ora Entin-Wohlman and Ben Svetitsky
for discussions.
This research is supported by the Israel Science Foundation under grant
222/02-1. MG is supported in part by the US Department of Energy.

\vspace{5ex}
\noindent {\large\bf Appendix A}
\secteq{A}
\vspace{3ex}

In this Appendix, we consider the issue of localization from
a somewhat different angle.  Instead of using our criterion
for localization of individual eigenmodes, Eq.~(\ref{lclz}),
we use Anderson's criterion of (partial) absence of diffusion
\cite{lcl}.   The argument reviewed here is originally due to
Ref.~\cite{ecco}, see also Ref.~\cite{DJT}.

In the language of condensed matter, the hermitian Wilson--Dirac operator
$H=H(m_0)$ is a ``tight-binding'' hamiltonian.  This means that
the ``electrons'' can reside only on the sites of a regular lattice.
This hamiltonian lives in four space dimensions, with
coordinates $x$ or $y$, and determines
the evolution of the system in a (fifth, continuous) time dimension,
with coordinate $s$.  We will assume that the electron encounters
a random potential on each lattice site, defined by some probability
distribution.  Since there is no back reaction of the electrons
on the random potential, this situation corresponds to the quenched
approximation of QCD, where of course the gauge fields
provide the random potential.
We introduce
the advanced Green function $\cg(x,y;s)$
and its Fourier transform $\cg(x,y;\l)$
\begin{eqnarray}
  \cg(x,y;s) &=&  \th(s) \sum_n \J_n(x) \J^\dagger_n(y) e^{-i\l_n s}\,,
\NON
  \cg(x,y;\l) &=& \sum_n \J_n(x) \J^\dagger_n(y) (\l -\l_n)^{-1} \,,
\label{green}
\end{eqnarray}
which solve the equations
\begin{eqnarray}
  (i\partial/\partial s - H) \cg(x,y;s) &=& i\d^4(x-y) \d(s) \,,
\NON
  (\l - H) \cg(x,y;\l) &=& \d^4(x-y) \,.
  \label{pdeq}
\end{eqnarray}
To carry out the Fourier transform, $\l$ has to be in the upper
half plane, whereas taking it in the lower half plane corresponds to
$\cg(x,y;\l)$ being the transform of the retarded Green function.

We now adopt the point of view that the electron is partially
localized if there is a non-zero probability for the electron
to be found at a specific location $y$ after an arbitrarily long time,
given that it started out at some other location $x$.
This probability is given by $\svev{|\cg(x,y;s)|^2}$ for
$s\to\infty$, where $\svev{\dots}$ denotes the
statistical average over the random potential at each lattice
site (or, in the QCD case, at each lattice link).
The idea is that this non-zero probability arises from
localized states, whereas the probability to escape to
infinity comes from extended states.  Obviously, in a finite
volume there is no clear distinction between these two
probabilities, but one may study the dependence of
$\svev{|\cg(x,y;s)|^2}$ for $s\to\infty$ on the volume.  If it
stays non-zero in the infinite-volume limit,
we have (partial) localization.

Following Ref.~\cite{ecco}, we may express the limiting value of
$\svev{|\cg(x,y;s)|^2}$ in terms of ensemble averages
of the Fourier transform $\cg(x,y;\l)$:
\begin{eqnarray}
  \lim_{s \to \infty} \svev{|\cg(x,y;s)|^2}
  &=&
  \lim_{\h \to 0} \h \int_0^\infty ds\,  e^{-\h s}\svev{|\cg(x,y;s)|^2}
\label{EC1} \\
  &=&
  \lim_{\h \to 0} \svev{
  \sum_{k,n} \J_n(x)\J_n^\dagger(y)\J_k(y) \J_k^\dagger(x)
   {\h\over \h + i(\l_n - \l_k)}}
\label{EC2} \\
  &=&
  \lim_{\h \to 0} {\h\over 2\p}
  \int d\l\, \svev{\cg(x,y;\l+i\h/2)\, \cg(y,x;\l-i\h/2)} \,.
\label{EC3}
\end{eqnarray}
Equality~(\ref{EC1}) follows after a partial integration (after
which the limit $\h\to 0$ can be taken); the other two follow
from elementary integrations. These equations are valid in finite
volume, and we will assume that they are valid in infinite volume as well.

We observe that the integrand on the right-hand side of Eq.~(\ref{EC3})
corresponds precisely to Eq.~(\ref{dblsum}),
\ie\ the pion two-point function, for $\l=0$ and identifying $\h/2=m_1$.
With the Anderson criterion for localization, we immediately
conclude that the $\l$-integral of the two-point function has a $1/\eta$
divergence provided $\lim_{s\to\infty}\svev{|\cg(x,y;s)|^2}>0$.
In finite volume this is always the case,
whereas in infinite volume, this is the case if the electron
is (partially) localized.

Anderson localization implies that, by definition,
the probability for the electron {\it not}
to escape to infinity, given by $\lim_{s\to\infty}\svev{|\cg(x,y;s)|^2}$,
is non-zero, and thus that
\begin{equation}
  \svev{\cg(x,y;\l+i\h)\, \cg(y,x;\l-i\h)}  \propto \h^{-1}\,,
    \qquad \h\to 0
\label{locc}
\end{equation}
for a range of $\l$
(note that this correlation function is positive).  It follows
that localization for a given $\l$ may be {\it defined} as the existence
of this divergence \cite{ecco}.  This coincides with our
analysis based on the mobility-edge hypothesis: if the mobility
edge $\l_c>0$, all modes with $\l<\l_c$ contribute to the
$1/m_1$ (or $1/\h$) divergence, and thus the right-hand side of
Eq.~(\ref{EC3}) does not vanish, leading to a non-zero probability
for the electron not to diffuse.

In reality, it may not be easy to measure the limiting value
of $\svev{|\cg(x,y;s)|^2}$ experimentally, because all the electronic
states are filled up to the Fermi energy.
The nature of the eigenstates close to the Fermi energy determine
the macroscopic properties (at sufficiently low temperature).
A disordered system is an insulator (zero electric conductivity)
if the electronic states at the Fermi energy are localized,
whereas a metallic behavior (non-zero conductivity) is observed
if the electronic states at the Fermi energy are extended.
A {\it metal-insulator} phase transition occurs when a mobility
edge reaches the Fermi energy. The Aoki phase transition is in this sense
a special kind of a metal-insulator transition.
For a discussion employing these concepts in the context of the chiral
phase transition in continuum QCD, see Ref.~\cite{chirdisorder}.

Finally, we mention that the concept of an effective lagrangian
for the long-range degrees of freedom which is used to study
the phase diagram of lattice QCD \cite{shsi,GSS}, is also widely used
in condensed-matter disordered systems, see Refs.~\cite{EF,KK}
and references therein.

\vspace{5ex}
\noindent {\large\bf Appendix B}
\secteq{B}
\vspace{3ex}

Using the operator formalism and the notation of Ref.~\cite{fs},
the numerator in Eq.~(\ref{mres}) is a sum of four positive terms.
For a fixed gauge field (and in an arbitrary normalization)
one of these terms is
\begin{equation}
  \left\langle 0' \left|
    \hatc_{\downarrow}(\vec{0},0)\, \hT^{N_s/2}\,
    \hatc^\dagger_{\downarrow}(\vec{x},\t)\,
    \hatc_{\uparrow}(\vec{x},\t)\, \hT^{N_s/2}\,
    \hatc^\dagger_{\uparrow}(\vec{0},0)
  \right|  0' \right\rangle \,.
\label{TOmRR}
\end{equation}
Here $\hT$ is the second-quantized transfer matrix.
$\hatc^\dagger$ and $\hatc$ are fermion creation and annihilation operators.
The up and down arrows represent two different flavors.
Spin indices have been suppressed.
The state $\sket{0'}$ is a reference state
annihilated by all the $\hatc$'s.
It encodes the boundary conditions in the fifth dimension.
(For notational simplicity we have given the result when the quark mass
is zero; all arguments generalize to the case of a non-zero mass,
as well as to the other three terms that we do not show.)

The composite operator
$\hatc^\dagger_{\downarrow}(\vec{x},\t)\, \hatc_{\uparrow}(\vec{x},\t)$
belongs to the pseudo-scalar density $J_{5q}^+(\vec{x},\t)$
located in the middle of the five-dimensional bulk. Therefore,
to be reached from one of the boundaries, one has
to act $N_s/2$ times with the transfer matrix
(we assume that $N_s$ is even).
If an admissibility condition is imposed,
the (first-quantized) transfer matrix $T(M,a_5)$ has a gap
(\ie\ there are no eigenvalues in some open neighborhood of one).
In the limit $N_s\to\infty$, $\hT^{N_s/2}$ becomes proportional to
the ground-state projector $\sket{0_H}\sbra{0_H}$.
(The ground state $\sket{0_H}$ of $\hT$ is obtained by filling the Dirac sea
of states that correspond to all eigenvalues $\o_n > 1$ of $T(M,a_5)$.)
In this case expression~(\ref{TOmRR}) is proportional to
$
\sbra{0_H}
  \hatc^\dagger_{\downarrow}(\vec{x},\t)\,
  \hatc_{\uparrow}(\vec{x},\t)
\sket{0_H}
$,
which is in fact identically zero, because the (in general non-zero) states
$\hatc_{\uparrow} \sket{0_H}$ and
$\hatc_{\downarrow} \sket{0_H}$ have different
flavors and, hence, are orthogonal. Thus, the numerator in Eq.~(\ref{mres}) vanishes
and chiral symmetry is recovered (exponentially)
in the limit $N_s\to\infty$. The case of realistic gauge actions
is discussed in Sect.~6.

\vspace{5ex}


\begin{thebibliography}{99}

\bibitem{wilson}
K.\ Wilson, in {\it New Phenomena in Sub-Nuclear Physics}
(Erice, 1975), ed. A.~Zichichi (Plenum, New York, 1977)

\bibitem{dwf}
D.\ B.\ Kaplan, \PLB{288} (1992) 342 [hep-lat/9206013];
Y.\ Shamir, \NPB{406} (1993) 90 [hep-lat/9303005]

\bibitem{fs}
V.\ Furman and Y.\ Shamir, \NPB{439} (1995) 54 [hep-lat/9405004]

\bibitem{oovlp}
R.\ Narayanan and H.\ Neuberger, \PLB{302} (1993) 62
[hep-lat/9212019]; \NPB{412} (1994) 574 [hep-lat/9307006];
\NPB{443} (1995) 305 [hep-th/9411108]

\bibitem{ovlp}
H.\ Neuberger, \PLB{417} (1998) 141 [hep-lat/9707022]

\bibitem{mlsymm}
P.H.\ Ginsparg, K.G.\ Wilson, \PRD{25} (1982) 2649
M.\ L\"uscher, \PLB{428} (1998) 342 [hep-lat/9802011]

\bibitem{morel}
A.\ Morel, J. Physique {\bf 48} (1987) 111

\bibitem{bg}
C.\ Bernard and M.\ Golterman, \PRD{46} (1992) 853 [hep-lat/9204007]

\bibitem{aoki}
S.\ Aoki, \PRD{30} (1984) 2653; {\bf 33} (1986) 2399; {\bf 34} (1986) 3170;
\PRL{57} (1986) 3136; \NPBP{60A} (1998) 206 [hep-lat/9707020]

\bibitem{bankscasher}
A.\ Casher, \PLB{83} (1979) 395;
T.\ Banks and A.\ Casher, \NPB{169} (1980) 103

\bibitem{creutz}
M. Creutz, \PRD{52} (1995) 2951; \RMP{73} (2001) 119 [hep-lat/0007032]

\bibitem{shsi}
S.\ Sharpe and R.\ Singleton, \PRD{58} (1998) 074501 [hep-lat/9804028]

\bibitem{KeSe}
R.\ Kenna and J.C.\ Sexton, \PRD{65} (2002) 014507

\bibitem{kawamotosmit}
N.\ Kawamoto and J.\ Smit, \NPB{192} (1981) 100

\bibitem{Aokiq}
S. Aoki and A. Gocksch, \PLB{231} (1989) 449; {\bf B243} (1990) 409;
\PRD{45} (1992) 3845;
S.\ Aoki, T.\ Kaneda and A.\ Ukawa, \PRD{56} (1997) 1808 [hep-lat/9612019]

\bibitem{Aokif}
S.\ Aoki, A.\ Ukawa and T.\ Umemura, \PRL{76} (1996) 873 [hep-lat/9508008];
\NPBP{47} (1996) 511 [hep-lat/9510014];
S.\ Aoki, T.\ Kaneda, A.\ Ukawa and T.\ Umemura, \NPBP{53} (1997) 438
[hep-lat/9612010];
K.\ Bitar, \NPBP{63} (1998) 829 [hep-lat/9709086]

\bibitem{EW}
M.B.\ Einhorn and J.\ Wudka, \PRD{67} (2003) 045004 [hep-ph/0205346]

\bibitem{SV}
J.\ Smit and J.\ Vink, \NPB{286} (1987) 485

\bibitem{excep}
W.\ Bardeen, A.\ Duncan, E.\ Eichten, G.\ Hockney
and H.\ Thacker, \PRD{57} (1998) 1633 [hep-lat/9705008];
W.\ Bardeen, A.\ Duncan, E.\ Eichten and H.\ Thacker,
\PRD{59} (1999) 014507 [hep-lat/9806002]

\bibitem{scri}
R.G.\ Edwards, U.M.\ Heller and R.\ Narayanan,
\NPB{522} (1998) 285 [hep-lat/9801015];
\NPB{535} (1998) 403 [hep-lat/9802016];
\PRD{60} (1999) 034502    [hep-lat/9901015]

\bibitem{jansenetal}
K.\ Jansen, C.\ Liu, H.\ Simma and D.\ Smith,
\NPBP{53} (1997) 262 [hep-lat/9608048]

\bibitem{GSS} M.\ Golterman, S.\ Sharpe and R.\ Singleton, in preparation

\bibitem{BNN}
F.\ Berruto, R.\ Narayanan and H.\ Neuberger, \PLB{489} (2000) 243
[hep-lat/0006030]

\bibitem{tm}
R.\ Frezzotti, P.A.\ Grassi, S.\ Sint and P.\ Weisz,
JHEP {\bf 0108} (2001) 058 [hep-lat/0101001];
R.\ Frezzotti, hep-lat/0210007

\bibitem{ks}
A.J.\ McKane and M.\ Stone, \APH{131} (1981) 36

\bibitem{aokidw} S.\ Aoki and Y.\ Taniguchi, \PRD{65} (2002) 074502
[hep-lat/0109022];
S.\ Aoki, \NPBP{109} (2002) 70 [hep-lat/0112006]

\bibitem{lcl}
P.W.\ Anderson, \PR{59} (1958) 1492;
N.F.\ Mott, J. Non-Cryst. Solids {\bf 1} (1968) 1

\bibitem{DJT}
D.J.\ Thouless, \PRP{13} (1974) 93

\bibitem{EF}
K.B.\ Efetov, Adv.\ Phys.\ {\bf 32} (1983) 53

\bibitem{KK}
B.\ Kramer and A.\ MacKinnon, Rep.\ Prog.\ Phys. {\bf 56} (1993) 1469

\bibitem{LS}
H.\ Leutwyler and A.\ Smilga, \PRD{46} (1992) 5607

\bibitem{GHK}
G.S.\ Guralnik, C.R.\ Hagen and T.W.B\ Kibble,
{\it Broken Symmetries and the Goldstone Theorem}, in
R.L. Cool and R.E. Marshak, Eds.,
{\it Advances in Particle Physics}, Vol. 2, Wiley, 1968

\bibitem{dwq} Y.\ Shamir, \PRD{59} (1999) 054506 [hep-lat/9807012]

\bibitem{pq}
C.\ Bernard and M.\ Golterman, \PRD{49} (1994) 486 [hep-lat/9306005];
O.\ B\"ar, G.\ Rupak and N.\ Shoresh, hep-lat/0210050

\bibitem{horvath}
I.\ Horvath, \PRL{81} (1998) 4063 [hep-lat/9808002]

\bibitem{hjl}
P.\ Hern\'andez, K.\ Jansen and M.\ L\"uscher,
\NPB{552} (1999) 363-378 [hep-lat/9808010]

\bibitem{dwpt}
S.\ Aoki, Y.\ Taniguchi, \PRD{59} (1999) 054510 [hep-lat/9711004];
P.\ Vranas, \PRD{57} (1998) 1415 [hep-lat/9705023];
S.~Aoki, T.~Izubuchi, Y.~Kuramashi and Y.~Taniguchi,
\PRD{\bf 59} (1999) 094505 [hep-lat/9810020];
S.~Aoki and Y.~Taniguchi, \PRD{\bf 59} (1999) 094506 [hep-lat/9811007]

\bibitem{yspt}
Y.\ Shamir, \PRD{62} (2000) 054513 [hep-lat/0003024]

\bibitem{effD}
H.\ Neuberger, \PRD{57} (1998) 5417 [hep-lat/9710089];
Y.\ Kikukawa and T.\ Noguchi, hep-lat/9902022

\bibitem{bor}
A.\ Borici, \NPBP{83} (2000) 771 [hep-lat/9909057];
hep-lat/9912040

\bibitem{rm}
P.\ Vranas, \PRD{57} (1998) 1415 [hep-lat/9705023];
T.\ Blum and A.\ Soni, \PRL{79} (1997) 3595 [hep-lat/9706023];
T.\ Blum, \NPBP{73} (1999) 167 [hep-lat/9810017];
S.\ Aoki, T.\ Izubuchi, Y.\ Kuramashi and Y.\ Taniguchi,
\NPBP{83-84} (2000) 624 [hep-lat/9909154];
S.\ Aoki, T.\ Izubuchi, Y.\ Kuramashi and Y.\ Taniguchi,
\PRD{62} (2000) 094502 [hep-lat/0004003];
T.\ Blum {\it et.\ al.} (RBC collaboration), hep-lat/0007038

\bibitem{Db}
C.\ Orginos (RBC collaboration) \NPBP{106} (2002) 721 [hep-lat/0110074];
Y.\ Aoki {\it et.\ al.} (RBC collaboration), hep-lat/0211023

\bibitem{sh}
S.\ Sharpe, \PRD{46} (1992) 3146 [hep-lat/9205020]

\bibitem{benetal}
R. C. Brower and B. Svetitsky, \PRD{61} (2000) 114511 [hep-lat/9912019];
F.\ Berruto, R.C.\ Brower and B.\ Svetitsky, \PRD{64} (2001) 114504
[hep-lat/0105016]

\bibitem{hopp}
M.\ Golterman and Y.\ Shamir, JHEP {\bf 0009} (2000) 006 [hep-lat/0007021]

\bibitem{PH}
P.\ Hern\'andez, \NPBP{106} (2002) 80 [hep-lat/0110218];
R.G.\ Edwards, \NPBP{106} (2002) 38 [hep-lat/0111009]

\bibitem{Iw}
A.\ Ali Khan {\it et al.} (CP-PACS collaboration),
\PRD{63} (2001) 114504 [hep-lat/0007014]

\bibitem{ftDW}
P.\ Chen {\it et.\ al.}, \PRD{64} (2001) 014503 [hep-lat/0006010];
P.\ Vranas, hep-lat/0001006

\bibitem{ecco}
E.N.\ Economou and M.H.\ Cohen, \PRL{25} (1970) 1445

\bibitem{chirdisorder}
R.A.\ Janik, M.A.\ Nowak, G.\ Papp and I.\ Zahed, \PRL{81} (1998) 264
[hep-ph/9803289]; \PLB{442} (1998) 300 [hep-ph/9807550]

\end{thebibliography}
\end{document}